\documentclass[a4paper,11pt]{article}
\usepackage{jinstpub}
\usepackage{lineno}
\usepackage[dvipsnames]{xcolor}
\usepackage{dcolumn}%
\usepackage{bm}%
\usepackage{natbib}
\usepackage{amsfonts}
\usepackage{amsmath}
\usepackage{siunitx}
\usepackage{graphicx} %
\usepackage{booktabs} %
\usepackage{tikz}
\usepackage{lipsum}
\usepackage{xargs} %
\usepackage[
    separate-uncertainty=true, 
    round-mode=uncertainty,
    round-precision=1,
]{siunitx} %
\usepackage[export]{adjustbox}  %
\usepackage{placeins}
\usepackage{hyperref}
\usepackage{subfiles}
\usepackage[normalem]{ulem}
\usepackage{multirow}
\usepackage{ragged2e}
\usepackage[colorinlistoftodos,prependcaption,textsize=tiny]{todonotes}
\usepackage{xspace}
\usepackage{orcidlink}
\usepackage{ulem}
\usepackage{booktabs}
\usepackage{xspace}

\definecolor{linkcolor}{HTML}{0b5394}

\renewcommand{\figurename}{Figure}
\newcommand{\cc}{\textsc{CaloHadronic}\xspace}

\newcommand{\dmodel}{\mbox{128}}
\newcommand{\halfdmodel}{\mbox{64} }

\newcommand{\ecalD}{\textsc{ECal edm-diffusion}\xspace}

\newcommand{\hcalD}{\textsc{HCal edm-diffusion}\xspace}

\newcommand{\pointsFM}{\textsc{PointCountFM}\xspace}

\newcommand{\Pandora}{\textsc{PandoraPFA}\xspace}

\hypersetup{
    colorlinks=true,
    linkcolor=linkcolor,     %
    citecolor=linkcolor,    %
    urlcolor=linkcolor       %
}

\title{\boldmath \cc : a diffusion model for the generation of hadronic showers}

\author[a,b]{Thorsten~Buss~\orcidlink{0000-0002-1717-2138},}
\author[b]{Frank~Gaede~\orcidlink{0000-0002-7055-9200},}
\author[a]{Gregor~Kasieczka~\orcidlink{0000-0003-3457-2755},}
\author[b]{Anatolii~Korol~\orcidlink{0000-0002-2569-1771},}
\author[b]{Katja~Kr\"uger~\orcidlink{0000-0002-1956-6608},}
\author[c]{Peter~McKeown~\orcidlink{0009-0006-9722-2233},} 
\author[a,1]{and Martina~Mozzanica~\orcidlink{0009-0002-1111-6247}\note{Corresponding author.}}
\emailAdd{martina.mozzanica@uni-hamburg.de}

\affiliation[a]{
	Institute for Experimental Physics, Universität Hamburg \\
	Luruper Chaussee 149, 22761 Hamburg, Germany
}

\affiliation[b]{
    Deutsches Elektronen-Synchrotron DESY, \\
    Notkestr. 85, 22607 Hamburg, Germany
}

\affiliation[c]{
    CERN, 1211 Geneva 23, Switzerland
}

\abstract{Simulating showers of particles in highly-granular calorimeters is a key frontier in the application of machine learning to particle physics. Achieving high accuracy and speed with generative machine learning models can enable them to augment traditional simulations and alleviate a major computing constraint. \\
Recent developments have shown how diffusion based generative shower simulation approaches that do not rely on a fixed structure, but instead generate geometry-independent point clouds, are very efficient. We present a transformer-based extension to previous architectures which were developed for simulating electromagnetic showers in the highly granular electromagnetic calorimeter of the International Large Detector, ILD. 
The attention mechanism now allows us to generate complex hadronic showers with more pronounced substructure across both the electromagnetic and hadronic calorimeters. This is the first time that machine learning methods are used to holistically generate showers across the electromagnetic and hadronic calorimeter in highly granular imaging calorimeter systems.
The code is available at \href{https://github.com/FLC-QU-hep/CaloHadronic}{https://github.com/FLC-QU-hep/CaloHadronic}.}


\arxivnumber{2506.21720}

\begin{document}

\maketitle
\flushbottom

\newpage
%\mm{ciao} \ak{ciao} \tb{ciao} \pem{ciao} \\
%\gk{ciao} \fg{ciao} \kk{ciao} 
\section{Introduction}

Building generative surrogates for expensive event generation and simulation tasks is a key step in enabling the physics program of the high-luminosity LHC (HL-LHC) and future collider studies~\cite{Apostolakis:2022nnf, Biedron:2022yst, Elvira:2022wyn}.
%Machine learning (ML) methods have been part of particle physics
%research for a long time.
%Over time, their applications have expanded widely, ranging from tasks like particle %tagging~\cite{Draguet:2912358,Karwowska:2024xqy,Mondal:2024nsa} and anomaly detection~\cite{ATLAS:2020iwa,ATLAS:2023azi,ATLAS:2023ixc,CMS:2024nsz} 
%to critical stages in reconstruction, such as particle tracking 
%~\cite{Burleson:2882507,ATL-PHYS-PUB-2024-018,Correia:2024kal} or even full event interpretation and %reconstruction~\cite{GarciaPardinas:2023pmx}.
%One significant application of ML in high-energy physics (HEP) lies in detector simulation. 

As experiments in high energy physics push the boundaries of luminosity resulting in ever increasing event rates, the computational demand of high-precision Monte Carlo (MC) simulations is growing to the point where it will soon surpass available computational resources~\cite{Adelmann:2022ozp}.
Generative models offer a promising solution to this challenge, potentially reducing the immense computational load required for these simulations. This has led to substantial research into the development of machine-learning architectures tailored for more efficient and accurate detector simulation~\cite{Butter:2022rso, Krause:2024avx}. Examples include generative adversarial networks (GANs)~\cite{Paganini:2017hrr, Paganini:2017dwg, deOliveira:2017rwa, Erdmann:2018kuh,
Musella:2018rdi, Erdmann:2018jxd, Belayneh:2019vyx, Butter:2020qhk, Javurkova:2021kms, Bieringer:2022cbs, Hashemi:2023ruu, 2024_atlas}, 
variational autoencoders (VAEs) and their variants
\cite{Buhmann:2020pmy, Buhmann:2021lxj, Buhmann:2021caf, 2024_atlas,Cresswell:2022tof,Diefenbacher:2023prl, hashemi2024deepgenerativemodelsultrahigh},
normalizing flows and various
types of diffusion \mbox{models ~\cite{sohldickstein2015deep,
song2020generative_estimatingGradients,
song2020improved_technieques_for_sorebased_geneerative, ho2020denoising,
song2021scorebased_generativemodelling, Mikuni:2022xry_CaloScore, Buhmann:2023bwk,
Acosta:2023zik, Mikuni:2023tqg,
Amram:2023onf, Chen:2021gdz, Krause:2021ilc, 
Krause:2021wez, schnake2022_pointFlow, Krause:2022jna,
Diefenbacher:2023prl, 
Xu:2023xdc, Buckley:2023daw,Ernst:2023qvn, OmanaKuttan:2024mwr,Favaro:2024rle, Brehmer:2024yqw}}, as well as generative pre-trained transformer (GPT) style models~\cite{Birk:2025wai}. The combination of a diffusion model with a transformer architecture, known as diffusion transformers~\cite{DBLP:attention, peebles23}, has been used in high-energy physics for jet generation~\cite{Leigh:2023pcjedi, Mikuni:2023fpcd, Butter:2025jetgpt, Mikuni:2025tar, Brehmer:2024yqw}.

The majority of these studies have focused on simulating electromagnetic showers, for a recent review see~\cite{HASHEMI2024100092}. Only a few of them (e.g.~\cite{Buhmann:2021caf}) have attempted to model hadronic showers, which exhibit significantly more complex and varied phenomenology with respect to electromagnetic ones. Additionally, studies have focused either on simulating showers in the electromagnetic calorimeter (ECal) or in the hadronic calorimeter (HCal) with high precision. 
In practice, however, a hadronic particle might already start interacting in the electromagnetic calorimeter and then continue showering in the hadronic calorimeter (and perhaps even beyond). 
Therefore, for realistic applications of these models, an efficient and comprehensive simulation of the entire shower is necessary.
This work represents the first approach to provide a unified framework for modeling hadronic showers across both the electromagnetic and hadronic components of a highly granular calorimeter system.

As electromagnetic and hadronic calorimeters typically have different cell sizes~\cite{DellaNegra:2012mga} --- leading to different spatial resolutions or, put differently, to a change in diameter of the shower when measured in cells (e.g.~\cite{Quast:2021dfg}) --- and are built from different materials (e.g.~\cite{ATLAScal}), correctly learning the behavior across the two detector systems is not trivial. Together with the much larger spatial extent and the much larger statistical fluctuations of hadronic showers compared to electromagnetic ones, this represents a major challenge for approaches based on regular grids.

The \cc model is a generative ML model able to generate showers in both ECal and HCal. The model uses a continuous normalizing flow (CNF)~\cite{Chen:2018wjc} for generating the number of points in each layer, coupled with EDM-diffusion models~\cite{karras2022edm} for simulating showers in both calorimeters. The model uses two separate EDM-diffusion models  (one for ECal and one for HCal), each designed to generate realistic showers based on their respective granularities. By using transformers within these models, \cc accurately generates showers in both calorimeters, respecting their distinct resolutions and materials. When generating a shower in the HCal, the ECal data is given as conditioning, and the attention mechanism captures the relevant transition between the two calorimeters.   

We demonstrate the holistic generation of showers for the planned International Large Detector at the International Linear Collider. Section~2 introduces the relevant detector components and data preparation steps. Next, Section~3 introduces the model architecture. We then first review the performance of the proposed approach at the level of simulated showers in Section~4 and also after passing through the standard reconstruction chain in Section~5. Section~6 concludes this work.

\section{Dataset}
The International Large Detector (ILD) \cite{ILD-IDR} is a detector concept proposed for future $e^{+}e^{-}$ Higgs factories, originally designed for operation at the International Linear Collider (ILC) \cite{ILCTDR2013}. The ILC is a proposed linear electron-positron collider that is supposed to initially operate at a center-of-mass energy of 250~GeV with the option to upgrade it to 1~TeV, with a rich program of Higgs-, Electroweak- and BSM physics \cite{ILCTDR2013}. 

The experimental community has developed designs for two complementary detectors, ILD and SiD, to optimally address the ILC goals. The emphasis for detectors at future Higgs factories is placed on ultimate precision. This requires detector technologies with new levels of performance. The momenta of the full set of final-state particles are best reconstructed with a Particle Flow Algorithm (PFA). This technique combines the information from the tracking system and from the calorimeter system to reconstruct the energy and the direction of all charged and neutral particles individually in the event. To minimize overlaps between neighboring particles, and to maximize the association accuracy between tracks and calorimeter clusters, calorimeters with very high granularity are needed. Therefore, both detector concepts employ highly granular calorimeters placed inside the solenoid coil and excellent tracking and vertexing systems. The two detector concepts differ in the choice of tracker technology. In this work, we focus on the ILD detector which would optimize the particle-flow resolution by making the detector large, thus better separating charged and neutral particles. \\
\paragraph{ECal}
Electromagnetic showers are measured with a compact highly-granular calorimeter with
absorber plates made of tungsten. The ECal barrel shape is octagonal with individual stacks laid out in a way to avoid projective dead zones in azimuth. The calorimeter is composed of 30 layers.
The sensitive medium consists of silicon sensors with about 5x5~mm$^{2}$ readout cells. \\
\paragraph{HCal} 
The hadronic calorimeter consists of 48 longitudinal sampling layers containing steel absorber plates. The active layers interleaved between the absorbers feature cells of size 3x3~cm$^2$, each comprising a scintillator tile readout individually by a silicon photomultiplier.\\
% {\kk{ leave out? already discussed above: The HCal layers are instrumented with high granularity for an efficient separation of 
% charged and neutral hadronic showers. This is necessary for particle flow, relying on minimal confusion between particles, as well as for a good muon identification.} \\
%for flavor jet tagging.  - muon ID is not relevant for flavor tagging\\

The ILD detector geometry is described using \textsc{DD4hep} \cite{dd4hep} version $1.30$, which provides an interface to the \textsc{Geant4} toolkit \cite{G4_toolkit} version $11.2.2$ for simulation\footnote{The \textsc{QGSP\_BERT} physics list is used in this study.}, as well as the standard suite of ILD reconstruction tools \cite{ILD-IDR}.

\paragraph{Data Generation}
The dataset \footnote{A subset of 20k example showers taken from the training set is available at \href{https://doi.org/10.5281/zenodo.15301636}{https://doi.org/10.5281/zenodo.15301636}.} used in this study was generated by firing single $\pi^{+}$ particles perpendicularly to the face of the ECal. $\pi^{+}$ particles were chosen, as they create hadronic showers with a distinct track like pattern resulting from ionization before the first nuclear interaction that is important to model correctly. The gun was positioned at $(x, y, z) = (-50~\text{mm}, 1804.7~\text{mm}, -150~\text{mm})$ in the ILD coordinate system. This coordinate system is defined such that the $z$-axis lies parallel to the beamline, the $y$-axis points vertically upward, and the $x$-axis completes the right-handed coordinate system. The particle gun was placed directly at the front surface of the ECal, so as to avoid interactions in the tracking system. The particles fly along the positive $Y$-axis, entering the calorimeter perpendicularly. The particle energies are uniformly distributed in the range of 10–90~GeV. A dedicated model of the ECal that has homogeneous sensitive layers without gaps between sensors was used during simulation, in order to avoid artifacts from such gaps when later placing the shower in a different position in the calorimeter.
% \tb{Shouldn't we also mention here that insensitive gaps in the layers ware removed to make the model applicable at (almost) any position of the detector? Also, I think the last sentence should be in presence. } \ak{Fixed the sentence tense. Regarding the regular detectors: It's hard to explain elegantly in a few senteces and I'm not sure we have to mention this here.} \fg{how about:  A dedicated model of the ECal that has homogeneous sensitive layers without gaps between sensors has been used here, in order to avoid artifacts from such gaps when later placing the shower in a different position in the calorimeter. ?}

\subsection{Data Preprocessing}
Following the approach in Ref.~\cite{Buhmann:2023bwk}, all \textsc{Geant4} steps 
are extracted from the sensitive layers of both calorimeters, yielding 2D point clouds for every one of the 
78 layers. Subsequently, each layer is projected onto a virtual grid with roughly 9 times higher granularity than the original ECal and HCal:
%different granularities for ECal and HCal. 
%The granularities of the grids are chosen 
%to reflect the resolutions of the calorimeter subsystems: a finer 
1.7x1.7~mm$^2$ granularity is used for the ECal, while a coarser granularity of 10x10~mm$^2$ is used for the HCal. Consequently, the total number of grid points per layer is $280,900$ for the ECal and $8,100$ for the HCal, given a cut along the transverse plane (square cut in the $x$-$z$ plane) from -450~mm to 450~mm. In a similar fashion to Ref. \cite{Buhmann:2023bwk} all layers are stacked along the $y$-axis, with the $y$-coordinate assigned according to the layer number. A random offset uniformly sampled from the interval $[-0.5,0.5]$ is applied to each point’s $y$-position, dequantizing the longitudinal shower profile.\\

To reduce the complexity of the point cloud and the total number of points, 
energy depositions at the Geant4 step level below the noise cutoff threshold $10^{-5}$~MeV are discarded. Given the pion incident energies in the range of 10-90~GeV, the coarser granularity point cloud has an average number of points around 1 700 and a maximum number of points of 5 000. Therefore, the final shape of one shower is [5 000, 4] with four being the number of features: $x$, $y$, $z$ and energy.

Standard scaling (i.e., zero mean and unit variance) is applied independently to the $(x, y, z)$ coordinates and to the logarithm of the energy deposition, $\log(E)$, as well as to the incident particle energy used for conditioning. This normalization ensures more stable training of the model and mitigates the effects of scale variation.

After model inference, the individual energy depositions produced by the model are positioned into the detector geometry using the \textsc{DDML}\footnote{\href{https://github.com/key4hep/DDML/}{https://github.com/key4hep/DDML/}} 
library~\cite{McKeown:2023QR, McKeown:2024jeq}, which uses the fast simulation hooks present in \textsc{Geant4} and \textsc{DD4hep}. This enables an efficient placement and scoring of hits in the sensitive elements of the detector.

% to write: energy cut at 10 to the minus 2, we use point clouds ..., 
%9x higher granularity..., 
% standardization for x,y,z and log energy and the energy standardization. 
% incident energy standardization and points per layer is point per layer divided by the max.

\section{Model and Architecture}
\label{section:architecture}
This section details the architecture of our proposed model, which is designed to simulate hadronic showers: \cc \footnote{The code for \cc including the hyperparamter settings used for training
is available on \href{https://github.com/FLC-QU-hep/CaloHadronic}{https://github.com/FLC-QU-hep/CaloHadronic}.}. The model is implemented using \textsc{PyTorch}~\cite{paszke2019pytorch}.

The \cc model for pion showers has three components. The first, \pointsFM, is a continuous normalizing flow
\cite{Chen:2018wjc} which is trained via flow matching \cite{lipman2023flow} to learn the number of points in the highly granular grid for each layer. The second and third components are both EDM diffusion \cite{karras2022edm} models which use a transformer mechanism \cite{DBLP:attention} for shower generation. The diffusion process has been implemented following~\cite{crowson2022kdiffusion}.
One of these diffusion models (\ecalD) is used to generate the ECal portion of the shower, while the other (\hcalD) generates the HCal portion. \\

The CNF with flow matching was chosen to generate the number of points 
per layer conditioned on the incident energy of the shower. 
Both features, points per layer in the respective calorimeters and incident energy, 
are then used in both the ECal and HCal diffusion models to generate the two parts of a shower. 
In \hcalD, the ECal part of the shower is also provided as an additional conditioning input. 
Pion showers are obtained by combining the outputs of the \ecalD and \hcalD models. 

\subsection{\texorpdfstring{\pointsFM}{PointCountFM}}
Normalizing flows~\cite{Rezende:2015ocs,Papamakarios:2019fms} consist of a diffeomorphism
between the physics and the latent space. The change of variables formula allows for direct
likelihood-based training. However, this requires the neural network to be invertible and
to have a tractable Jacobian, which imposes architectural constraints. Continuous
normalizing flows~\cite{Chen:2018wjc} extend this idea by modeling the
transformation as a neural ordinary differential equation (ODE) integrated over a time
variable. This approach allows us to choose almost any neural network architecture.
However, evaluating the likelihood now requires solving the ODE at each training step,
significantly slowing down the training process.

Flow matching~\cite{Albergo:2022iol,lipman2023flow} offers an efficient
alternative. Instead of solving the ODE, it constructs a time-dependent velocity field.
This allows training using a simple mean squared error loss, enabling training with a
single forward and backward pass. Score-based diffusion models using a probability flow
ODE can be seen as a special case of flow matching models.

\pointsFM is a continuous normalizing flow trained with conditional flow
matching~\cite{lipman2023flow}. It generates the number of points deposited in
each of the 78 calorimeter layers conditioned on the incident energy. For the network
architecture, we use a multi-layer perceptron with five hidden layers. We apply a series
of preprocessing steps to prepare the data: dequantization to handle discrete inputs, a
logarithmic transformation to stretch the high-density regions, and standard scaling to
normalize the features for stable training. All further hyper-parameters can be found in
Table~\ref{tab:pointsFM_config}.

We generate new samples from the trained model with the Heun second-order ODE solver using 200 steps.

\subsection{ECal and HCal blocks}
\label{subsec:impr}
Our model builds on two core ideas popularized by CaloClouds2 \cite{Buhmann:2023kdg}: the use of EDM diffusion for generative modeling and representing showers as point clouds. While our approach is inspired by these principles, we propose a distinct architecture tailored to hadronic showers, introducing new components in the model structure. 

Diffusion models \cite{pmlr-v139-nichol21a} have become a powerful approach for 
modeling data distributions. 
They operate by progressively corrupting data through a forward process and then learning a reverse denoising process to recover the original distribution, 
effectively establishing a diffusion-based transformation between structured data and a simple prior distribution. Score-Based Continuous-Time 
Discrete Diffusion Models \cite{sun2023scorebasedcontinuoustimediscretediffusion} extend discrete diffusion models by formulating the process in continuous time. 
The key difference is that they explicitly model the score function, which represents the gradient of the log-likelihood at time $t$. 
This allows for greater flexibility in sampling and training, enabling the use of stochastic differential equations (SDEs) or 
ordinary differential equations (ODEs) for efficient inference. They also allow custom choice of the number of steps for generation and distillation 
for even faster inference. \\
EDM diffusion \cite{karras2022edm} improves the reverse process of diffusion models by using more accurate numerical methods (higher-order solvers) 
that stabilize the generation process. It reduces the number of steps required for high-quality samples and speeds up inference leading 
to faster and more reliable diffusion sampling.

The new components in the model structure include:

\paragraph{\textsc{Monotonic weighting}}  
It was shown in \cite{kingma2023understanding} that all commonly used diffusion 
model objectives equate to a weighted integral of evidence lower bound (ELBOs) over different noise levels, 
where the weighting depends on the specific objective used. In the paper it is proved that 
if the weighting $w(\lambda _{t})$ is monotonic, then the weighted diffusion objective of
\[
\mathcal{L}_w(x) = \frac{1}{2} \, \mathbb{E}_{t \sim \mathcal{U}(0,1),\, \epsilon \sim \mathcal{N}(0, I)} \left[ 
w(\lambda_t) \cdot \left( -\frac{d\lambda}{dt} \cdot \left\| \epsilon_\theta(z_t; \lambda_t) - \epsilon \right\|_2^2 \right) 
\right]
\]
is equivalent to the ELBO with data augmentation (additive noise). 

The key insight is that the weighting function $w(\lambda)$ can be chosen to be monotonic in order to give a bit more weighting to
lower noise levels, which is where the model is most accurate. \\

Inspired by the non-monotonic EDM weighting function of EDM diffusion 
\cite{karras2022edm}, they evaluate a variant referred to as \textit{EDM-monotonic}. 
This function matches the original EDM weighting \( \tilde{w}(\lambda) \), 
except it is made monotonic by setting  
\( w(\lambda) = \max_\lambda \tilde{w}(\lambda) \) 
for all \( \lambda < \arg\max_\lambda \tilde{w}(\lambda) \).  

\paragraph{\textsc{Fourier Layer}}
Ref.~\cite{rahaman2019on} found empirical evidence of a spectral bias: i.e. lower frequencies are learned first in a neural network 
and learning higher frequencies gets easier with increasing manifold complexity. They show that mapping the inputs to a higher dimensional 
space using high frequency functions before passing them to the network enables better fitting of data that contains high frequency variation. 
Ref.~\cite{DBLP:journals/corr/abs-2003-08934} then leveraged these findings in the context of neural scene representations. \\
In the same way, one could enhance the sensitivity of the model to higher frequency by manually embedding the input.
This is done by applying a composition of two functions to the input. Using the same formalism as \cite{DBLP:journals/corr/abs-2003-08934}, 
we can use the composition $F_{\Theta} \; o \; \lambda$ where $F_{\Theta}$ is the learned while $\lambda$ is not . $F_{\Theta}$ 
is a regular MLP while $\lambda$ is a mapping $\mathbb{R} \rightarrow \mathbb{R}^{2L}$ defined by:
\begin{equation}
\lambda(p) = [\sin(2^{0} \pi p), \; \cos(2^{0} \pi p), \; \dots, \; 
            \sin(2^{L-1} \pi p), \; \cos(2^{L-1} \pi p)]. 
\end{equation}
In our case L is set to 10.

\paragraph{\textsc{Attention}} 
Hadronic showers have multiple differences with respect to electromagnetic showers. One of them is the common appearance of
tracks within the shower. These tracks are due to charged secondary particles traversing multiple calorimeter layers without undergoing an inelastic collision or being stopped. In the previous work of 
CaloClouds II \cite{Buhmann:2023kdg}, the authors used a
point-wise approach to model electromagnetic (EM) 
showers. Since EM showers are topologically more uniform and 
isotropic (typically with an ellipsoid-like shape), the point-wise 
layers were sufficient to model 
the structure of EM showers while adding very little computational overhead. However in the case of 
$\pi^{+}$ showers, the point-wise approach fails to capture 
complex interactions between points, such as tracks, as it treats
each point independently. This is problematic 
for hadronic showers, where modeling the fine structure of the shower is crucial. The attention mechanism, and 
consequently the transformer architecture \cite{DBLP:attention}, 
offers a potential solution to this problem.
Attention allows the relationships 
between different points at different positions 
in the shower to be modeled, capturing the 
underlying structure and improving the fidelity of generated hadronic shower events.\\ 

\begin{figure}[htbp]
\centering
\includegraphics[width=150mm]{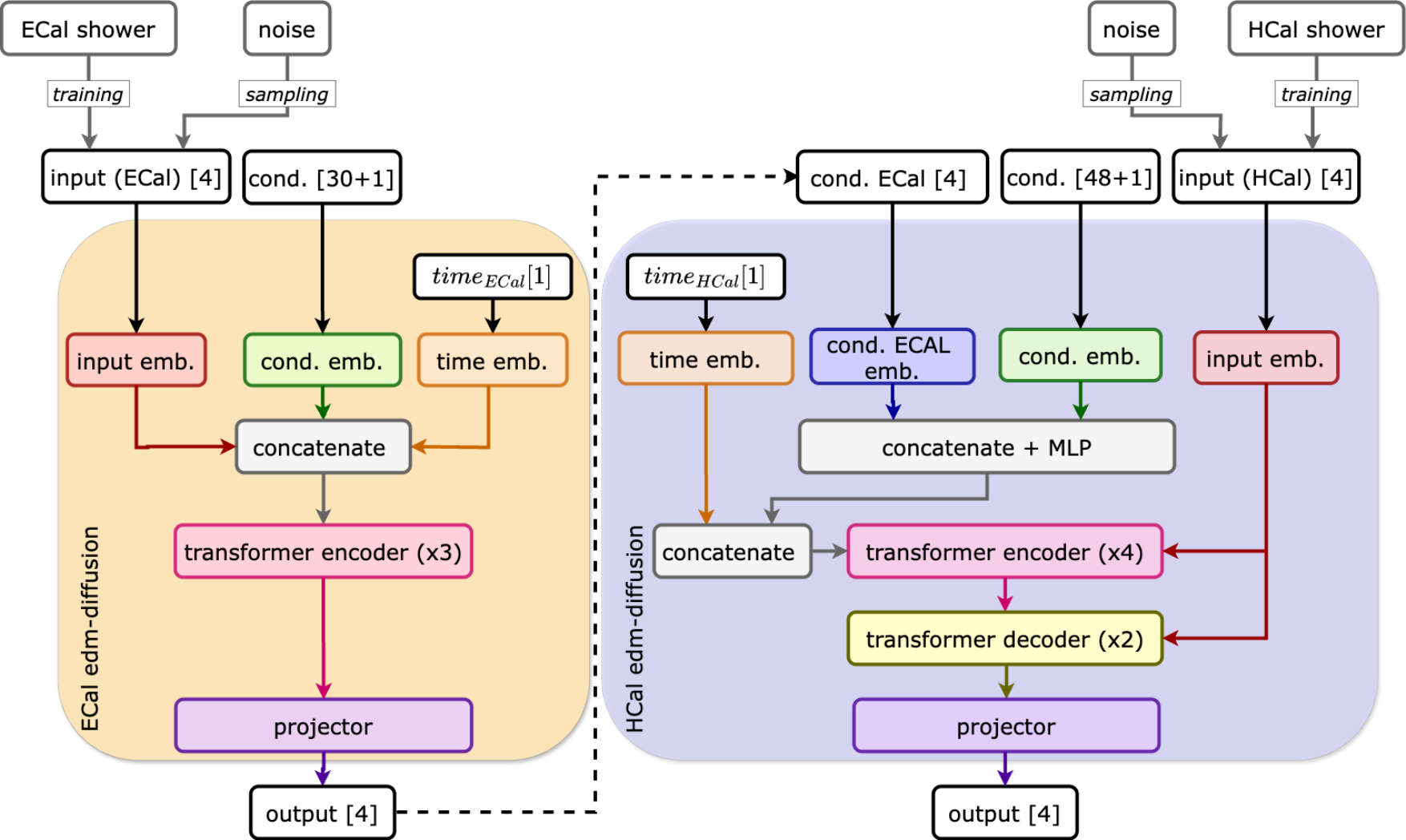} 
\caption{Illustration of the structure of the \ecalD (left) and \hcalD (right). 
$time$ is the time of the diffusion model used to add noise to the data and $cond.$ 
are the incident energy and the number of points per layer. The electromagnetic calorimeter (ECal) consists of 30 layers, and the hadronic calorimeter (HCal) consists of 48 layers. 
\textit{input (ECal or HCal)} is the input shower to be learned during training.}
\label{fig:diag_ecal_hcal}
\end{figure}

An illustration of the two blocks that compose the model --- \ecalD
%(for the ECal part)
and \hcalD 
%(for the HCal part) 
--- is shown in figure \ref{fig:diag_ecal_hcal}. Both blocks are based on EDM diffusion. 

\paragraph{\ecalD}\label{paragraph:ecal}
An illustration of \ecalD is shown on the left of figure \ref{fig:diag_ecal_hcal}.
For \ecalD, the conditional features are the incident energy 
and the number of points per layer for each of the 30 layers of the electromagnetic calorimeter of ILD. 
In their embedding, both of them are concatenated and then passed through 
two fully connected layers. The time feature of the diffusion model also has 
an embedding, consisting of a Fourier feature mapping and then two fully connected layers. 
The Fourier mapping of the time variable is taken from \cite{karras2022edm} and is 
similar to the previously mentioned Fourier Layer. The 4D input, consisting of the 
three coordinates ($x$,$y$ and $z$) and the energy ($e$) of the input shower, goes into custom 
defined embedding layers based on the Fourier Layer and then two 
fully connected layers as well. Time, conditioning and input embeddings, with a dimension 
of \dmodel, are concatenated along the points axis and fed into three layers of 
transformer encoder \cite{DBLP:attention}. After removing the conditioning points, 
a projection with three fully connected layers is applied to map to the desired output dimension of four features: $x$, $y$, $z$, $energy$. 

\paragraph{\hcalD}\label{paragraph:hcal}
An illustration of \hcalD is shown in figure \ref{fig:diag_ecal_hcal} on the right. 
In \hcalD the conditional features are the incident energy, 
the number of points per layer and the ECal shower (all points). The hadronic calorimeter of ILD consists of 48 layers. 
The time feature of the diffusion model has an identical embedding as explained 
in \ecalD . This is also true of the incident energy and the number of points per layer, 
with the only exception being the embedding dimension \halfdmodel instead 
of \dmodel. The ECal conditioning is first passed 
into a set compression layer. Drawing inspiration from \cite{Ji2024}, this layer maps the point cloud data to a fixed-size set of latent tokens, 10 in this case. More information on this layer can be found in appendix \ref{sec:appSCL}.
The ECal showers are then fed into two fully connected layers, concatenated together with the conditioning vector along the points axis and then passed through one linear layer to match the model dimension of \dmodel. At this point it is concatenated with the time feature. \\
The input of the model, which is the HCal part of the pion shower, 
has the same embedding as the input of \ecalD \ref{paragraph:ecal}. 
Both the input and the conditional features are fed into four layers of 
transformer encoder \cite{DBLP:attention}. Its output is then concatenated with the input and passed through two layers of transformer decoder \cite{DBLP:attention}. 
Finally, a projection consisting of three fully connected layers is applied to produce
the desired output dimensionality. \\

\begin{figure}[htbp]
    \centering
    \includegraphics[width=140mm]{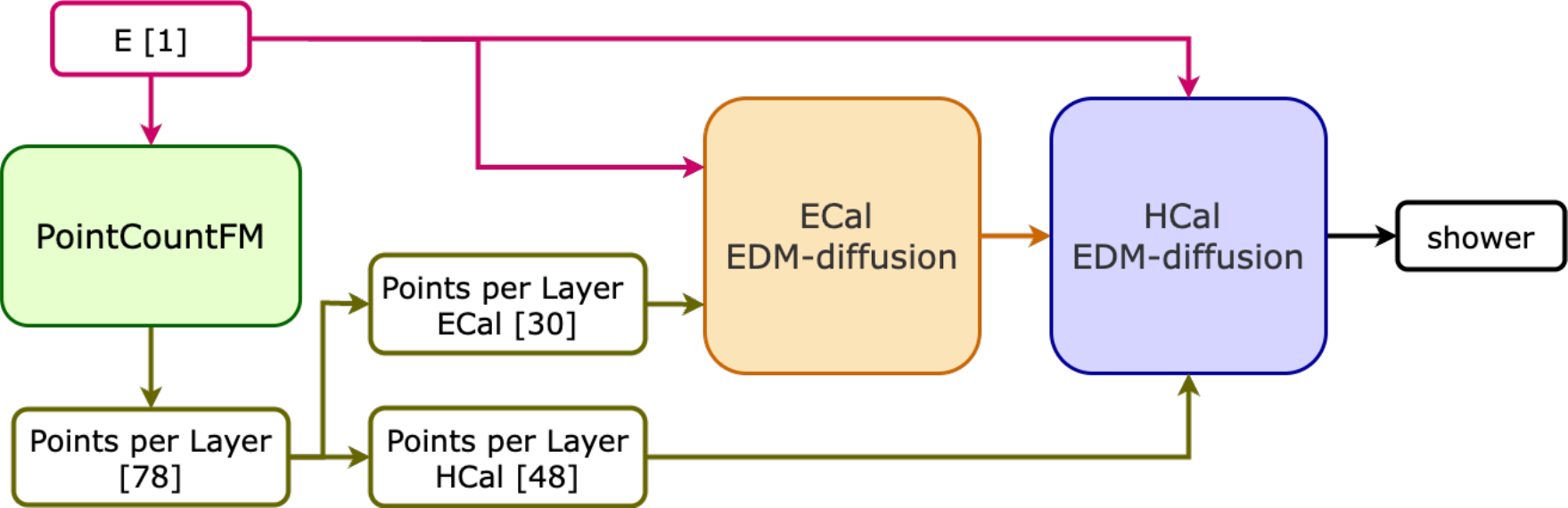}
    \caption{Illustration of the generation process of \cc. 
    The \pointsFM generates the number of points per layer. 
    The ECal points per layer distributions along with E are fed into the 
    \ecalD block. Its output, E and the HCal points per layer distributions 
    are passed into the \hcalD block. By concatenating the output of 
    the two blocks a new pion shower is generated.}
    \label{fig:diag_tot}
\end{figure}

\paragraph{Training and Sampling}\label{paragraph:sampling}
During training a random continuous time step is added to the input
and the model is trained to denoise the data.
The EDM diffusion models are trained conditioned on the 
shower energy and number of points per layer. The loss 
is approximated by a simple mean squared error (MSE) between 
a rescaled version of the input data and the denoised output.\\
During sampling, the conditional 
\pointsFM generates the number of points per layer for a 
given incident energy. The EDM diffusion blocks start from 
gaussian noise and use a stochastic sampler that combines 
a 2nd order deterministic ODE integrator with explicit 
Langevin-like “churn” — a process of periodically adding and removing noise to encourage exploration and improve sample diversity. During sampling, 
the inputs in figure \ref{fig:diag_ecal_hcal} 
start as Gaussian noise and the points are then moved step-by-step to create a shower. The Number of Function Evaluations (NFE) refers to the number of times a neural network that is used to model the reverse diffusion process needs to be evaluated during the sampling phase to produce a final sample. Since \cc is a continuous time score-based diffusion model, one has the flexibility to choose the NFE by finding a good compromise between precision and speed-up. In \cc 59 NFE were used.
An illustration of shower generation is shown in figure \ref{fig:diag_tot}. \\

The new components in the model training process include:
\paragraph{\textsc{Optimizer}} 
The Adam \cite{kingma2017adammethodstochasticoptimization} optimizer 
is a popular choice for training deep learning models due to its adaptive 
learning rate and momentum features.
However, it can be memory-intensive, especially when training large models
such as diffusion models.
In \cite{zhang2025adammini}, the authors propose a new optimizer called
Adam-mini, which is a variant of Adam that uses a smaller memory footprint
and is more efficient for training diffusion models.
The idea for Adam-mini can be understood by delving into the Hessian structure 
of neural networks and recalling the fact that it is a near-block diagonal matrix. 
In \cite{zhang2025adammini}, the authors find that, for each of these dense sub-blocks,
there exists a single high-quality learning rate that outperforms Adam. The memory footprint 
can be reduced by assigning a single learning rate to each of these dense sub-blocks 
instead of a separate learning rate per parameter as in Adam. 

\paragraph{\textsc{Scheduler}} 
Drawing inspiration from \cite{rombach2022highresolutionimagesynthesislatent}, the learning 
rate scheduler has been switched from using a constant learning rate with linear decay 
to OneCycleLR to improve training stability and convergence. 
OneCycleLR starts with a linear warm-up, allowing the model to gradually increase the 
learning rate, which helps stabilize the initial phase of training. 
After the warm-up, the scheduler applies cosine annealing decay, 
enabling faster convergence and avoiding premature reduction of the learning rate 
during fine-tuning. In \cite{smith2017cyclicallearningratestraining} 
\cite{smith2018superconvergencefasttrainingneural} it has been shown that OneCycleLR 
accelerates convergence and improves generalization, making it particularly effective 
for complex models like transformers and diffusion models.

\section{Results}
\begin{figure}[h!]
   \centering 
   \includegraphics[width=75mm]{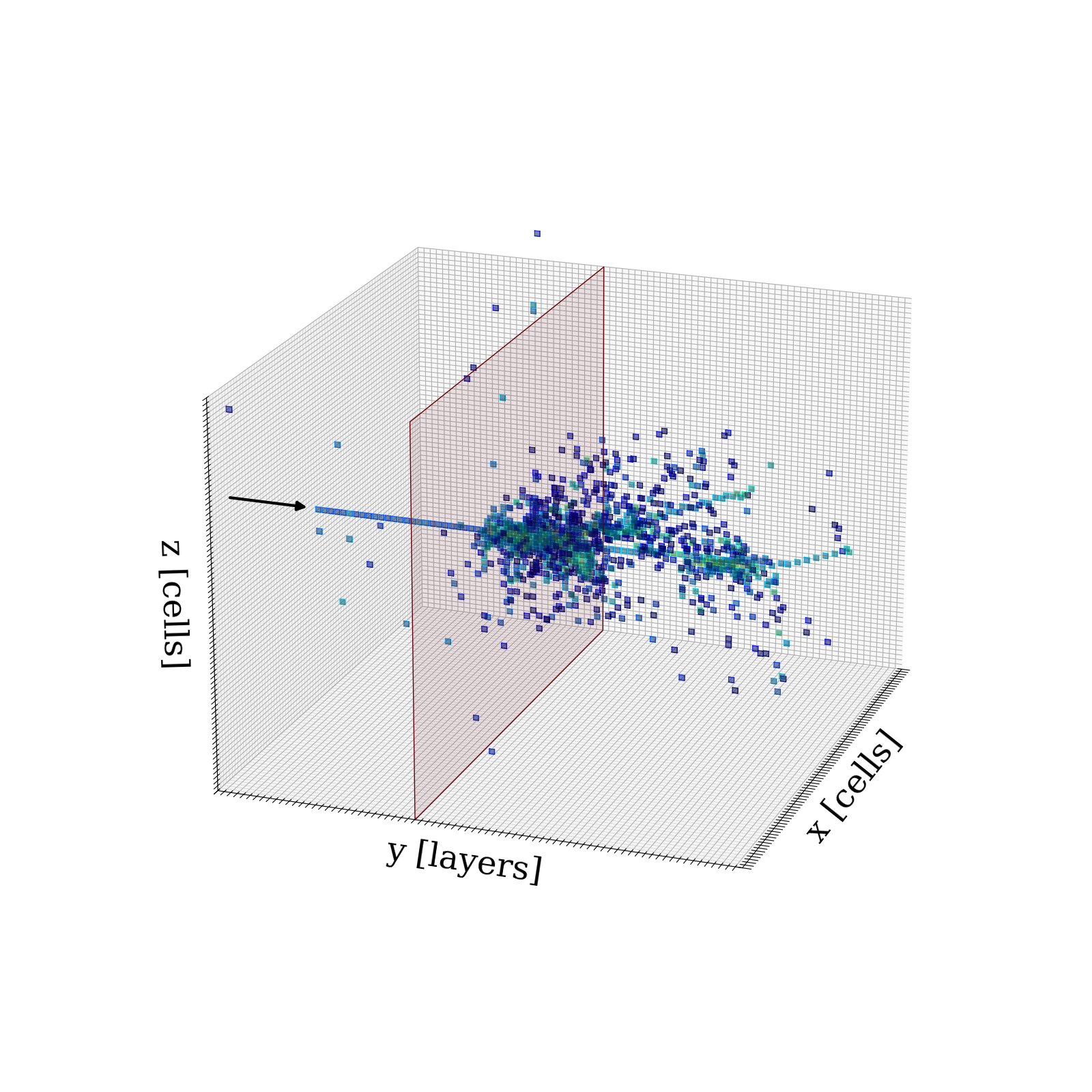}
   \includegraphics[width=75mm]{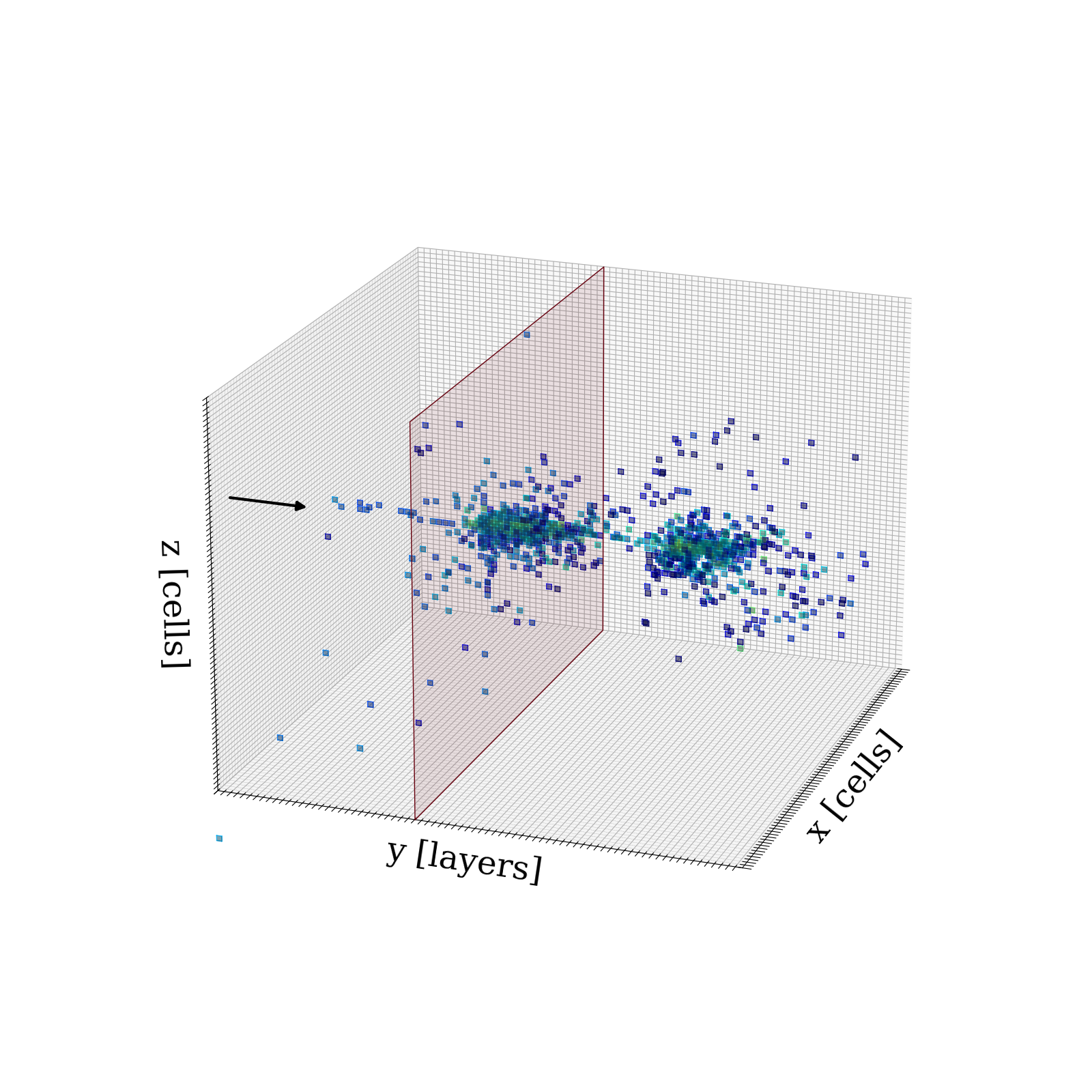}
   \caption{3D view of a 50~GeV $\pi^{+}$ shower simulated with Geant4 (left) 
   and a 50~GeV shower generated with \cc (right). The color represents the energy
   deposition in the cells. The red plane represents the division between ECal and HCal at layer 30.}
   \label{fig:3d}
\end{figure}

In the following, the results of generating  $\pi^{+}$ showers 
in the electromagnetic and hadronic calorimeter are presented.
After the sampling procedure explained in \ref{paragraph:sampling}, the 
effects of preprocessing were reversed during post-processing and a 
calibration on the number of points per layer was applied. Although not strictly necessary —since the model also predicts the $y$-axis— this calibration is used because the \pointsFM model better captures the distribution of points per layer, whereas the diffusion model can result in a less precise approximation along the layer ($y$) coordinate.
% While in CaloClouds2 there were multiple calibration, i.e. point per layer, 
% energy per layer and center of gravity along two cell axis, here resulted only
% in a single calibration. This one being the point per layer calibration. 
The calibration was done 
by considering the number of points per layer computed by \pointsFM and then assigning the 
generated points by \cc to the corresponding layer. \\ 
The final point cloud showers are projected back to a 
regular grid in which the cell size is the same as that 
of the ILD calorimeters, hence 5x5~mm$^{2}$ for the ECal and 
30x30~mm$^{2}$ for the HCal. For downstream analyses a cell 
energy cut at $\sim$$10^{-2}$~MeV is applied, since below this 
threshold the sensor response is indistinguishable from 
electronic noise. This cut was applied to all showers when 
calculating the shower observables and scores in this section. 
A first naive comparison of the showers generated by \cc with Geant4 simulated $\pi^{+}$ 
showers can be made from a visual inspection of whether the final projection of the showers 
resembles the Geant4 equivalent. In figure \ref{fig:3d} the 3D view of a
50~GeV Geant4 shower (left) and a 50~GeV \cc shower (right) are shown. 

Figure \ref{fig:3d} shows that the tracks resulting from hadronic interactions occurring in $\pi^{+}$ showers, are well reproduced by \cc. This was possible due to the self-attention mechanism that took care of the interactions among points.
The shower core is well defined and the tracks are visible. Notably, the CaloHadronic model better reproduces fine-grained, track-like structures in the HCal, which are especially evident when compared to the Geant4 reference. Noise manifests as low-energy deposits spread outside the core of the shower and can arise from detector effects such as electronics noise, cross-talk, or background activity. More generated
$\pi^{+}$ showers can be found in appendix \ref{sec:appC}. 

\subsection{Physics Performance}\label{subsec:pp}
\begin{figure}[h!]
   \centering 
   %\resizebox{\textwidth}{!}{
   \includegraphics[width=47.5mm]{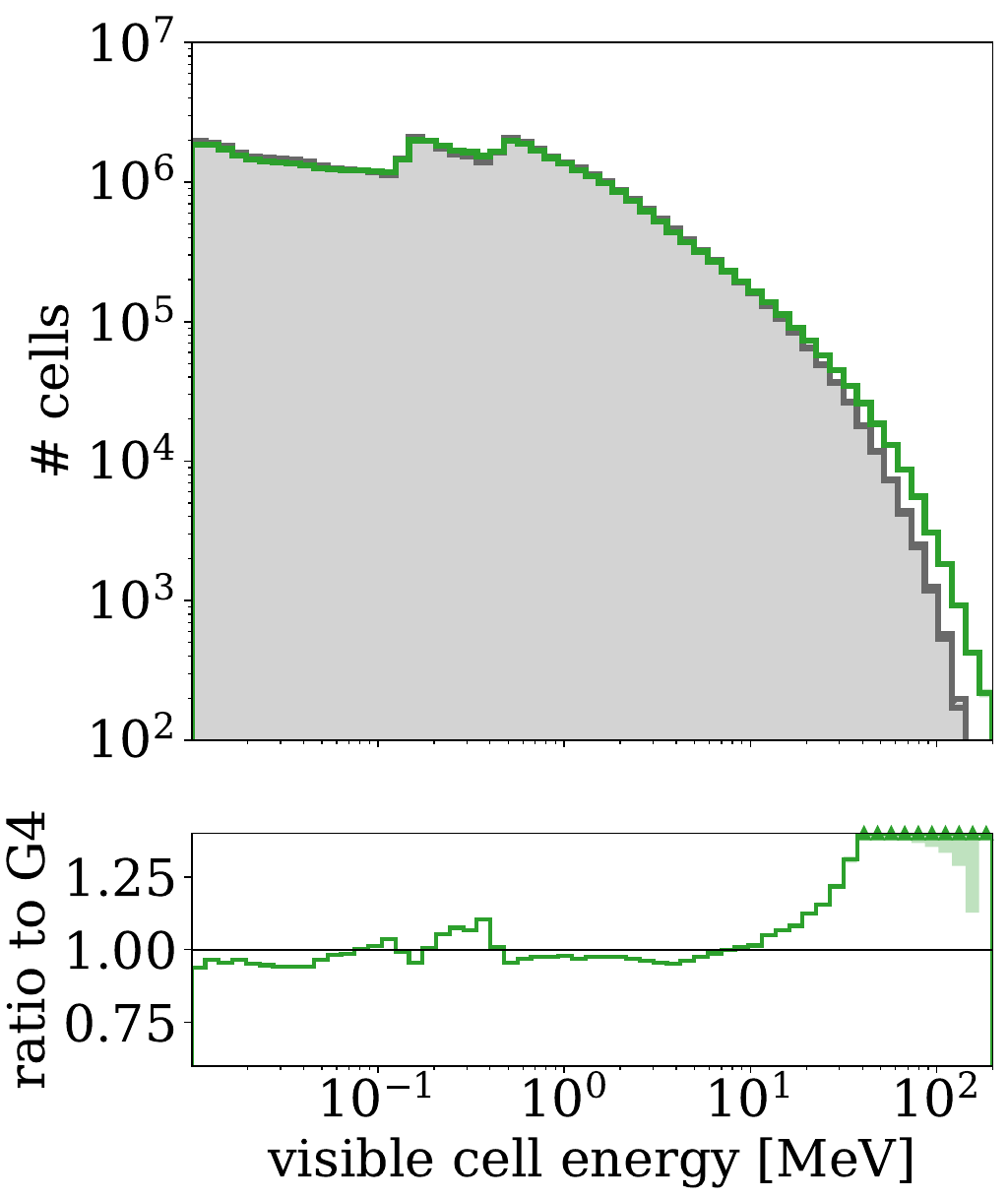}
   \includegraphics[width=44.7mm]{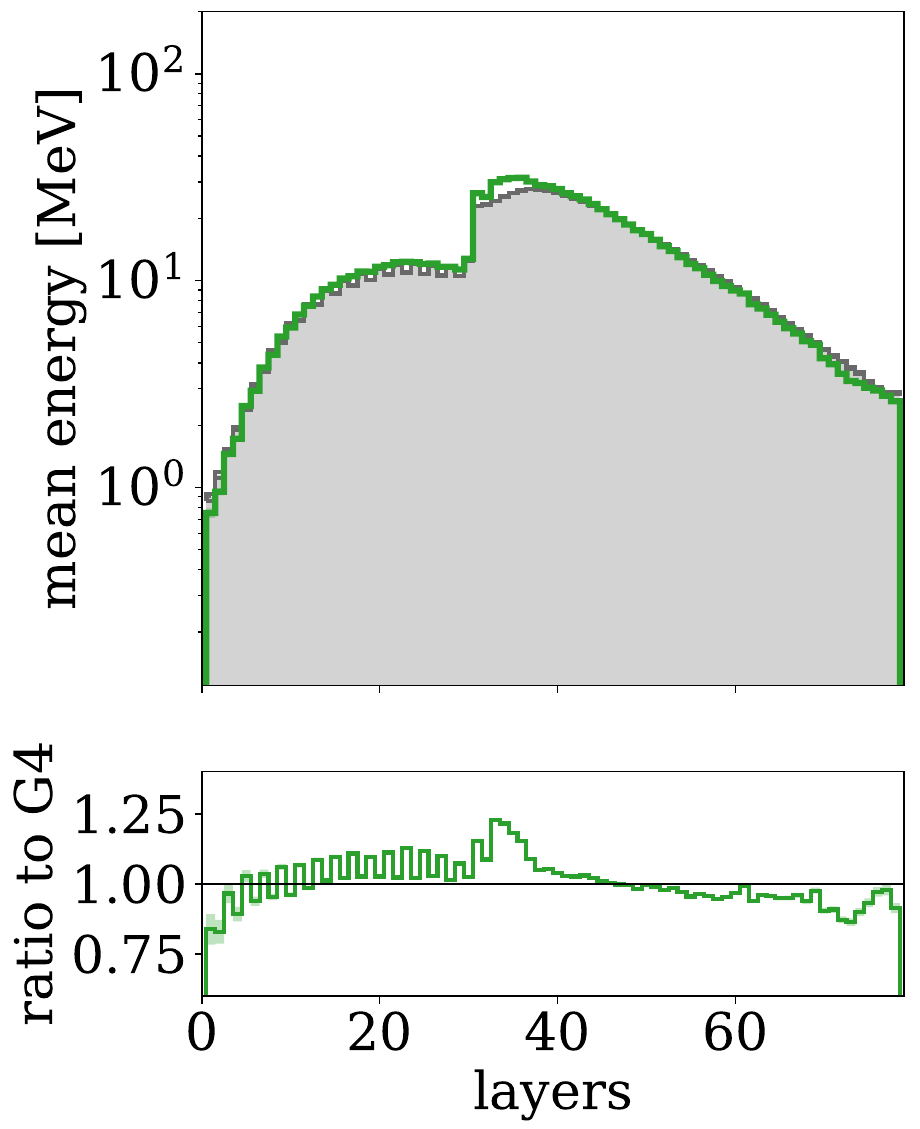}
   \includegraphics[width=44mm]{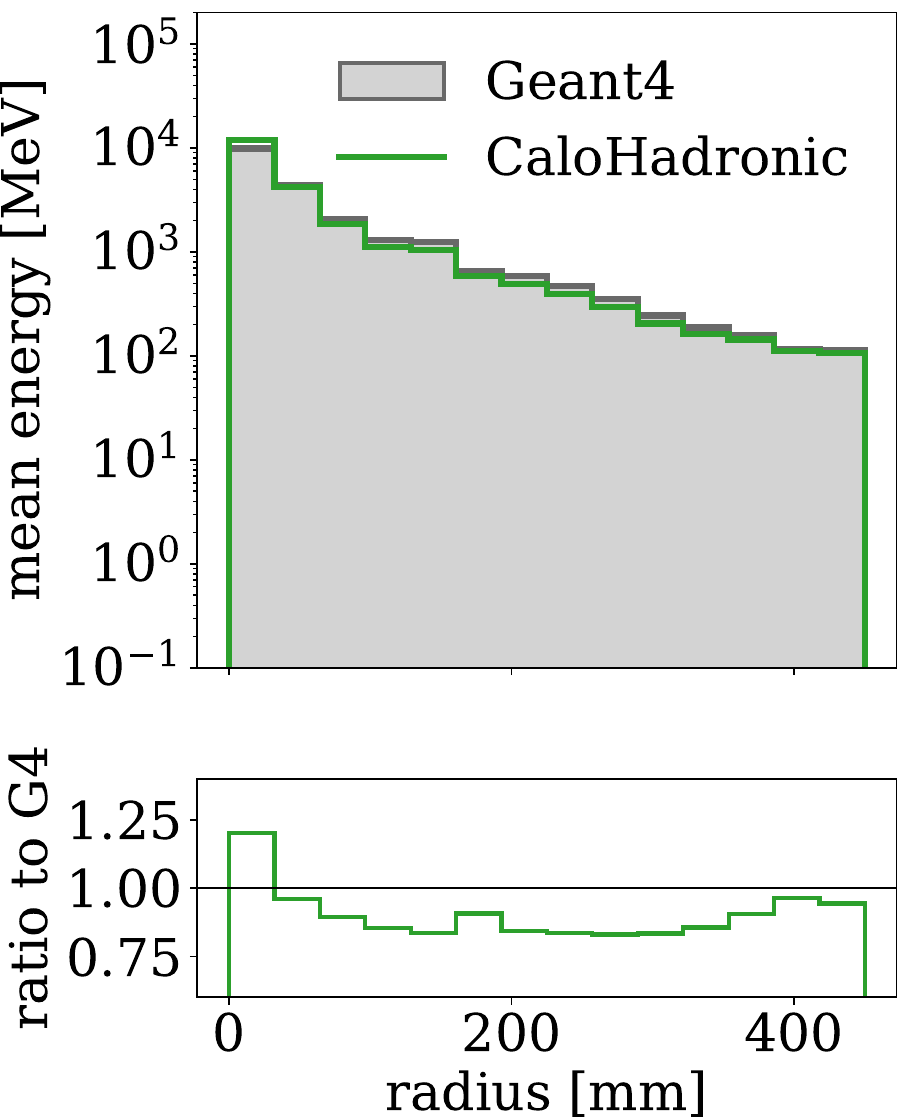}
   %}
   \caption{Histogram of the cell energies (left), 
   longitudinal shower profile (center), and radial shower profile 
   (right) for Geant4 and \cc. All distributions are calculated with 
   $50 000$ events sampled with a uniform distribution of incident 
   particle energies between 10 and 90 GeV. The bottom panel provides 
   the ratio to Geant4. The error band corresponds to the 
   statistical uncertainty in each bin.}
   \label{fig:dist1}
\end{figure}

\begin{figure*}
   \centering 
   \hspace*{-2cm}
   \resizebox{\textwidth}{!}{
   \includegraphics[width=150mm]{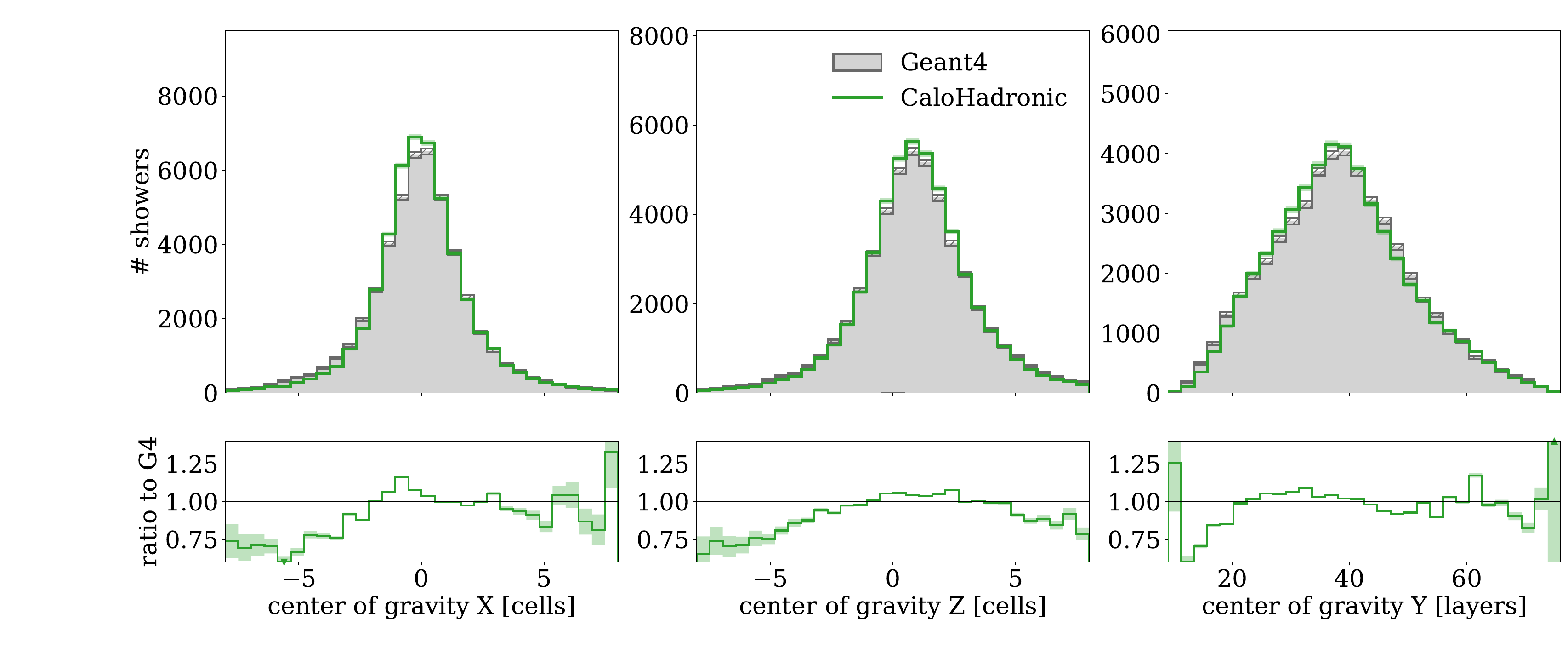}
   }
   \caption{Position of the center of gravity of showers along the $x$ (left), 
   $z$ (center), and $y$ (right) directions. All distributions are calculated for $50 000$ showers with a uniform distribution of incident particle energies 
   between 10 and 90 GeV. The error band corresponds to the statistical uncertainty 
   in each bin.}
   \label{fig:dist2}
\end{figure*}

In this section, various calorimeter shower distributions 
for both the Geant4 test set and datasets generated using 
\cc ~are compared. In figures \ref{fig:dist1} and \ref{fig:dist2} 
cell-level distributions are shown for 50k $\pi^{+}$ showers across both the 
ECal and HCal with an incident energy uniformly distributed 
between 10-90~GeV.\\ 

In figure \ref{fig:dist1} the energy distribution per cell (left), as well as the longitudinal (center) and radial (right) shower profiles 
are shown. 

\begin{figure*}
   \centering 
   \includegraphics[width=150mm]{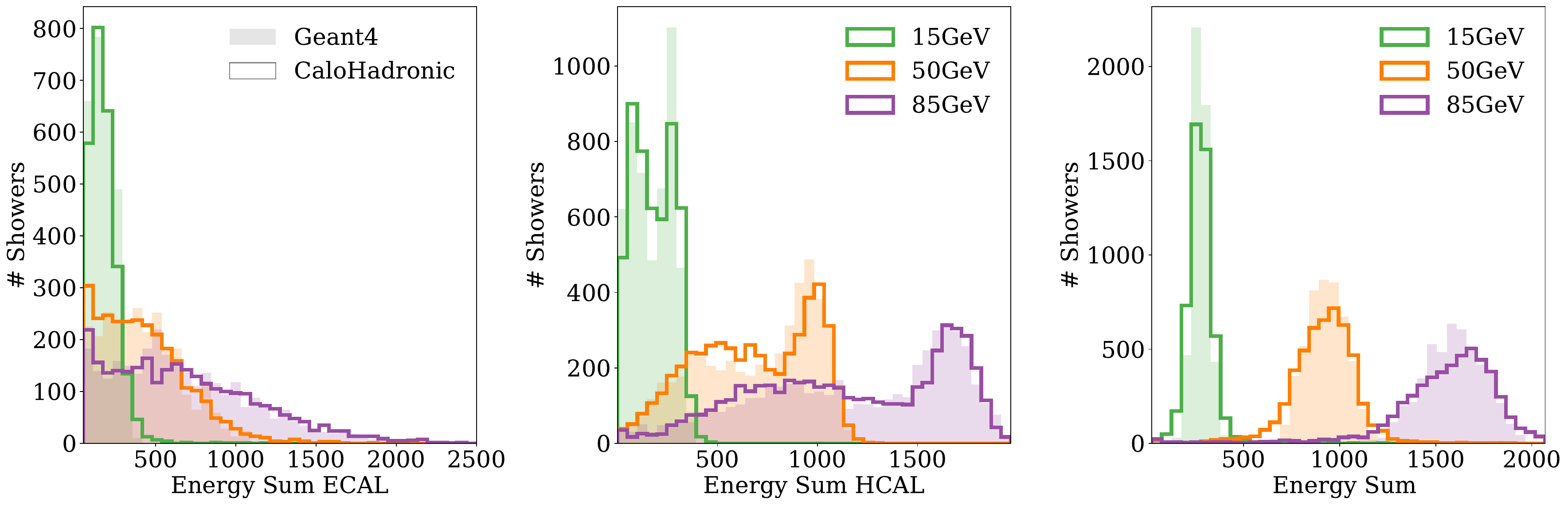}
   \caption{Total visible energy per shower at incident energies of 15~GeV (green), 
   50~GeV (orange), and 85~GeV (purple) for ECal (left), HCal (center), and for
   both calorimeters combined (right). For each energy, 5 000 showers are shown. Geant4 showers are sampled within $\pm$ 1 MeV of the target energy; \cc showers are newly generated at the same incident energies.}
   \label{fig:ensum}
\end{figure*}

The per-cell energy distribution (figure \ref{fig:dist1}, left) 
contains the energy of the cells of all showers in the test
dataset. In this distribution there are two peaks. 
The first one at $\sim$0.2~MeV corresponds to the most probable 
energy deposition of a minimum ionizing particle (MIP) in the 
silicon sensors of the ECal while the second one at $\sim$1~MeV 
corresponds to the MIP of the HCal. The model reproduces both peaks and the overall shape of the distribution reasonably well. The largest deviations are confined to the high-energy tail, where the number of contributing cells is small and the statistical fluctuations are correspondingly larger. The excess of energy could be due to the electromagnetic subsystem of the detector, since the ECal has a smaller cell size. This could affect the total energy distribution. However, the results of the total visible energy presented in figure \ref{fig:ensum} show that the effect of this overestimation is small.

The longitudinal shower profile (figure \ref{fig:dist1}, center) 
describes how much energy is deposited on average in each of the 
78 calorimeter layers. The first 30 layers are part of the ECal, while the 
last 48 compose the HCal. The model describes the distribution well. 
It can be observed that there are small bumps in the ECal layers, 
where the Geant4 simulation differs between even and odd layers. 
This alternating structure of higher and lower energy depositions per layer arises from the fact that for technical reasons, pairs of silicon sensors 
are placed on either side of one tungsten absorber layer facing opposite directions. 
These slabs are then installed into a tungsten structure containing every other absorber layer.
The different amount of nonsensitive material results in the 
observed pair-wise difference in the mean energy
between consecutive layers. While Geant4 reproduces this effect, \cc shows noticeable difficulties in modeling the subtle alternating pattern. This reflects a small limitation in capturing layer-dependent shower modulations. However these mismodelings only have a small effect on reconstruction for hadronic showers, as it is mostly a limitation for electromagnetic showers for which there are alternative generative models.

The radial shower profile (figure \ref{fig:dist1}, right) 
describes the average radial energy distribution around the central
shower axis (in the $y$-direction) in both ECal and HCal. \cc has a tendency to generate slightly narrower showers than Geant4. The shower profile is presented in appendix \ref{sec:appD} in figure \ref{fig:distxyz}. Although these deviations are small, the slightly narrower showers produced by the model compared to Geant4 could reduce the separation power of particle flow object (PFO) reconstruction, particularly in environments where accurate discrimination between overlapping showers is critical. Nevertheless, \cc shows overall very good agreement with Geant4. \\

In figure \ref{fig:dist2} distributions for the center of gravity 
(the energy weighted shower centroid, CoG) in the $x$ (left), 
$z$ (center), and $y$ (right) directions are shown. 
\cc models the CoG in the $x$,
$z$ and $y$ directions relatively well. The CoG in $x$ and
$z$ are slightly too narrow, while the CoG along $y$ is slightly shifted to the left. 
It should be noted that the center of gravity along the 
incident direction ($y$) is also affected by the points per 
layer calibration of \pointsFM.

We now investigate the performance of \cc at 
specific incident energies. Given that the dataset contains 
energies ranging from 10 to 90~GeV, incident energies of 15~GeV, 
50~GeV, and 85~GeV were selected to represent the lower, 
middle, and upper ends of the range, respectively. The Geant4 incident energies are selected by extracting 5,000 showers from the initial dataset within a $\pm$ 1 MeV window centered on the chosen incident energies. The \cc showers, in contrast, are generated anew using the same incident energies as those used for Geant4. \\ 

\begin{figure*}
   \centering 
   \includegraphics[width=150mm]{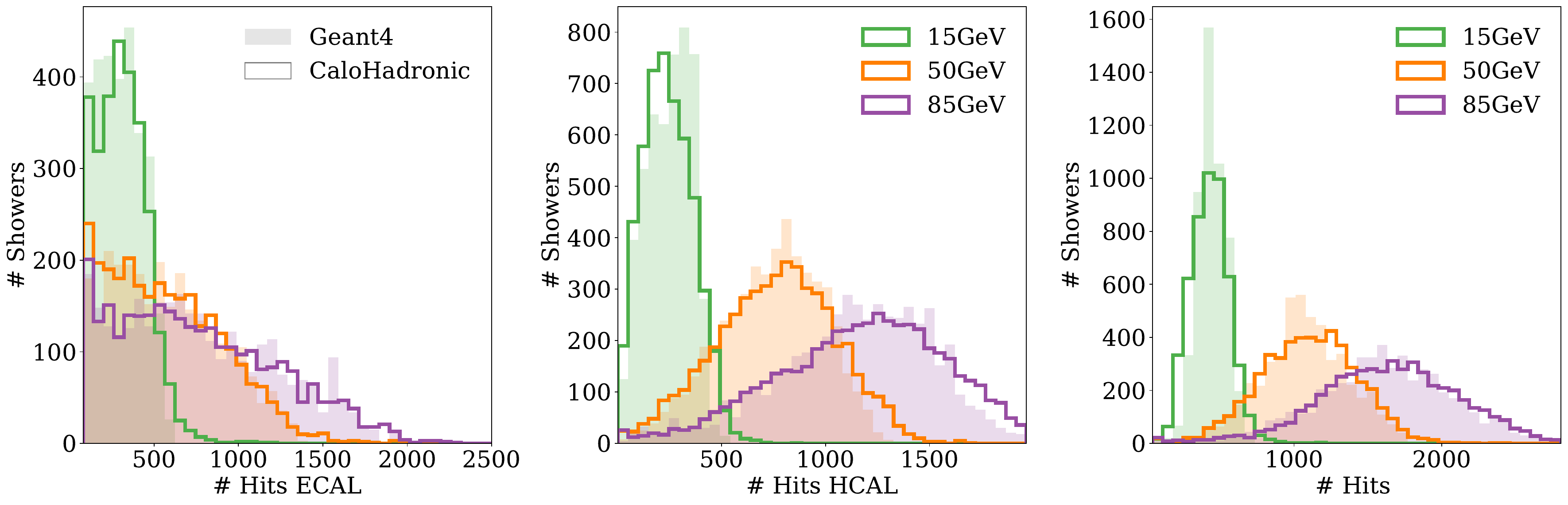}
   \caption{Number of hits per shower at incident energies of 15~GeV (green), 
   50~GeV (orange), and 85~GeV (purple) for ECal (left), HCal (center), and for
   both calorimeters combined (right). For each energy, $5,000$ showers are shown. Geant4 showers are sampled within $\pm$ 1 MeV of the target energy; \cc showers are newly generated at the same incident energies.}
   \label{fig:hits}
\end{figure*}

In figure \ref{fig:hits} the number of hits for 5k showers at three
incident energies is shown for the ECal and HCal individually, as well as for both calorimeters combined.
The number of hits is defined as the number of cells with energy
above the threshold of 0.02~MeV.
In figure \ref{fig:ensum} the energy sum (total visible energy) for 5k showers at three incident energies for ECal, HCal and both calorimeters combined is shown. As noted earlier in the cell energy spectrum shown in Figure \ref{fig:dist1}, \cc tends to slightly overestimate the hit energy, which could in principle affect the total energy distribution. However, the results here show that the impact of this overestimation is very minor.
The total energy and the number of hits are represented well by \cc. In appendix \ref{app_res} both resolution and linearity plots are also shown.

\subsection{Correlation Studies}\label{subsec:correlation}

In this section, 2D histograms between the shower-level observables described in \ref{subsec:pp} are studied. These histograms enable an investigation into how well some correlations are learned by {\cc}. Investigations are performed considering both the entire incident 
energy range (10-90~GeV) and for single incident energies. \\

\begin{figure*}
   \centering 
   \includegraphics[width=150mm]{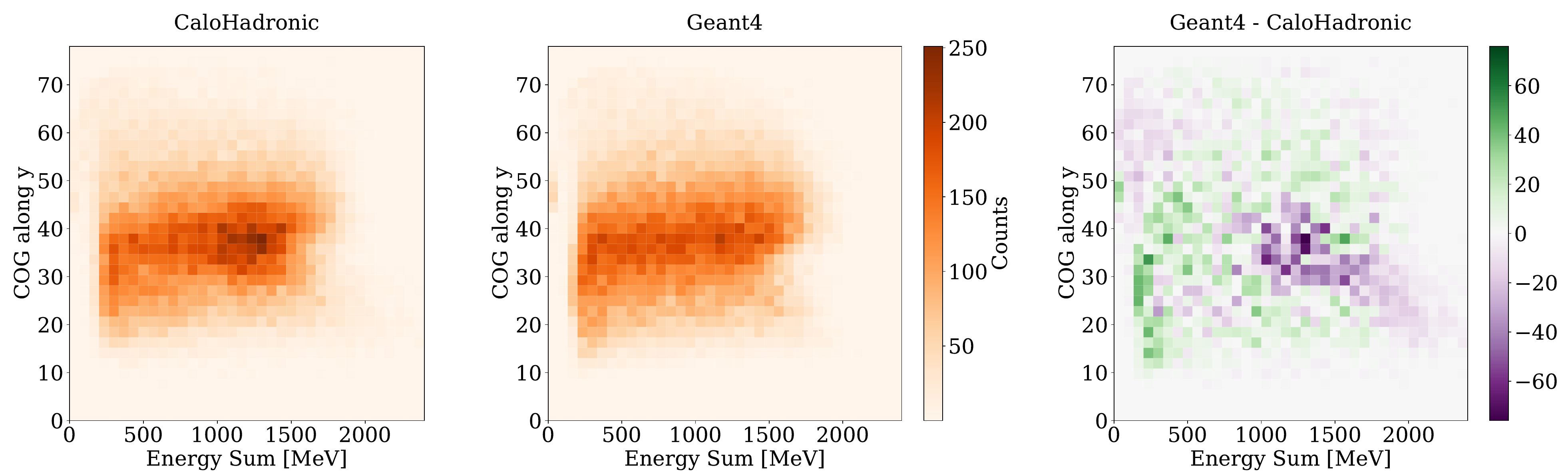}
   \caption{2D histograms of the center of gravity in the $y$ direction and the energy sum per shower for 50k showers with incident energies in the range of 10-90~GeV. The left plot shows the \cc distribution, the center plot shows the 
   Geant4 distribution. The color represents the number of showers. The right histogram shows the difference between the two.}
   \label{fig:corrCOGY}
\end{figure*}

In figure \ref{fig:corrCOGY} 2D histograms of the center of gravity versus the
energy sum per shower are shown for the incident energy range of 10-90~GeV for 50k showers. 
% \kk{I'm a bit puzzled by this plot. I would expect to see a logarithmic dependence of the GoGy on the incident energy, and therefore also on the Energy sum. Not sure I see this here (neither for Geant nor for our model)} \mm{we could discuss this in a meeting} 
The difference between the two 2D histograms is also shown, in order to highlight regions with significant deviations while accounting for statistical fluctuations. 

\begin{figure*}
   \centering 
   \includegraphics[width=150mm]{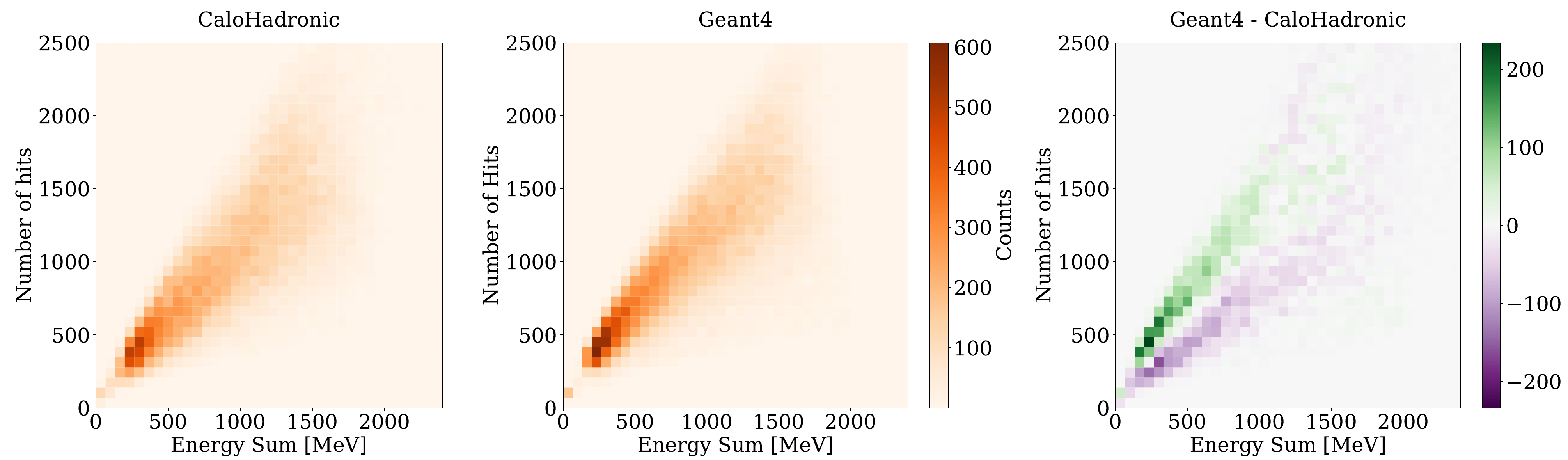}
   \caption{2D histograms of the number of hits and the energy sum per shower for 50k showers with incident energies in the range of 10-90~GeV. 
   The left plot shows the \cc distribution, the center plot shows the 
   Geant4 distribution. The color represents the number of showers. The right histogram shows the difference between the two.}
   \label{fig:corrN}
\end{figure*}

In figure \ref{fig:corrN} 2D histograms between the number of hits and the
energy sum per shower, as well as their difference, are shown for the incident energy range of 10-90~GeV for 50k showers. \\

In figures \ref{fig:corrCOGY} and \ref{fig:corrN} both histograms show good agreement between \cc and Geant4. Deviations are visible if one looks at the difference. 
In figure \ref{fig:corrCOGY} the region where the CoG along y is between the 30th and 50th layer exhibits the largest discrepancies between Geant4 and \cc . Within this range, \cc tends to over-populate the central part of the energy sum distribution while under-populating the tails. The difference in the remaining regions of the histogram can be considered negligible, as the statistics are relatively low (fewer than 50 showers per bin).
In figure \ref{fig:corrN}, the orange 2D histograms show a good agreement between Geant4 and \cc distributions. However, the difference plot reveals a systematic shift. For a fixed number of hits, \cc shows an excess of events at lower energy sums and a deficit at higher ones, compared to Geant4. On the other hand, for a fixed energy sum, \cc has fewer events at lower hit counts and more at higher hit counts.

\begin{figure*}
   \centering 
   \includegraphics[width=155mm]{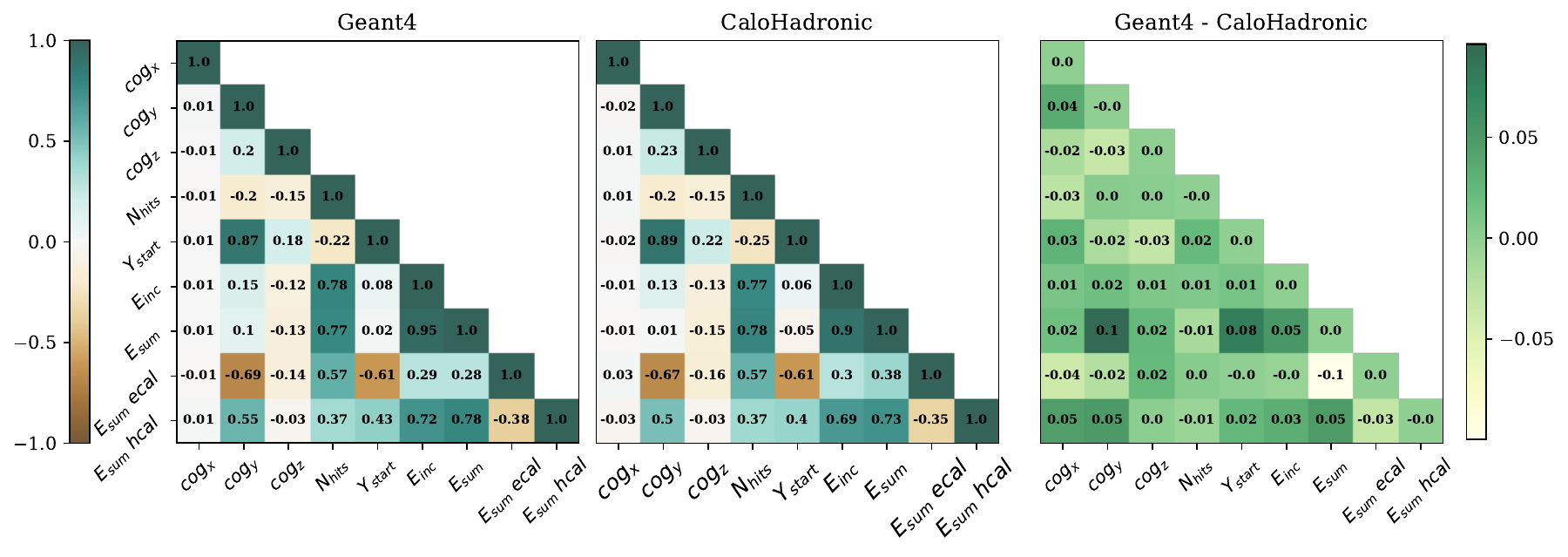}
   \caption{Pearson correlation coefficient matrix for key shower observables for both Geant4 (left) and \cc (center). The rightmost panel displays the difference between the two matrices.}
   \label{fig:pearson}
\end{figure*}

To investigate the correlation aspects of $\pi^{+}$ shower generation in greater depth, one can compute the Pearson correlation coefficient between relevant observables, such as the three center of gravity observables, incident energy, shower start layer, total visible energy and total visible energy in the ECal and HCal separately per shower. In the absence of precise timing information, we define the "shower start layer" as the layer containing the most energetic cell. While this does not guarantee that the shower physically starts in this layer, it provides a practical proxy based on the assumption that the initial stages of the shower typically deposit the largest energy near the start.
This coefficient quantifies the linear relationship between two variables, providing a value between -1 and 1, where 1 indicates a perfect positive correlation, -1 a perfect negative correlation, and 0 no linear correlation. The table for the Pearson correlation coefficient for Geant4 and \cc is presented in figure \ref{fig:pearson} along with their difference. 

Figure \ref{fig:pearson} shows a good agreement between the Pearson correlation coefficient of Geant4 and \cc. As one can see from the rightmost table, the variables in which \cc coefficients vary most significantly with respect to Geant4 (Pearson coefficient difference > 0.07) are the ones with total visible energy versus the center of gravity along y, shower start layer, and the total visible energy deposited in the ECal. 

% \begin{figure*}
%    \centering 
%    \includegraphics[width=160mm]{figures/results/Correlation_histograms_10-90GeV.png}
%    \caption{correlations 10-90GeV.}
%    \label{fig:corr10-90}
% \end{figure*}

\begin{figure*}
   \centering 
   \includegraphics[width=150mm]{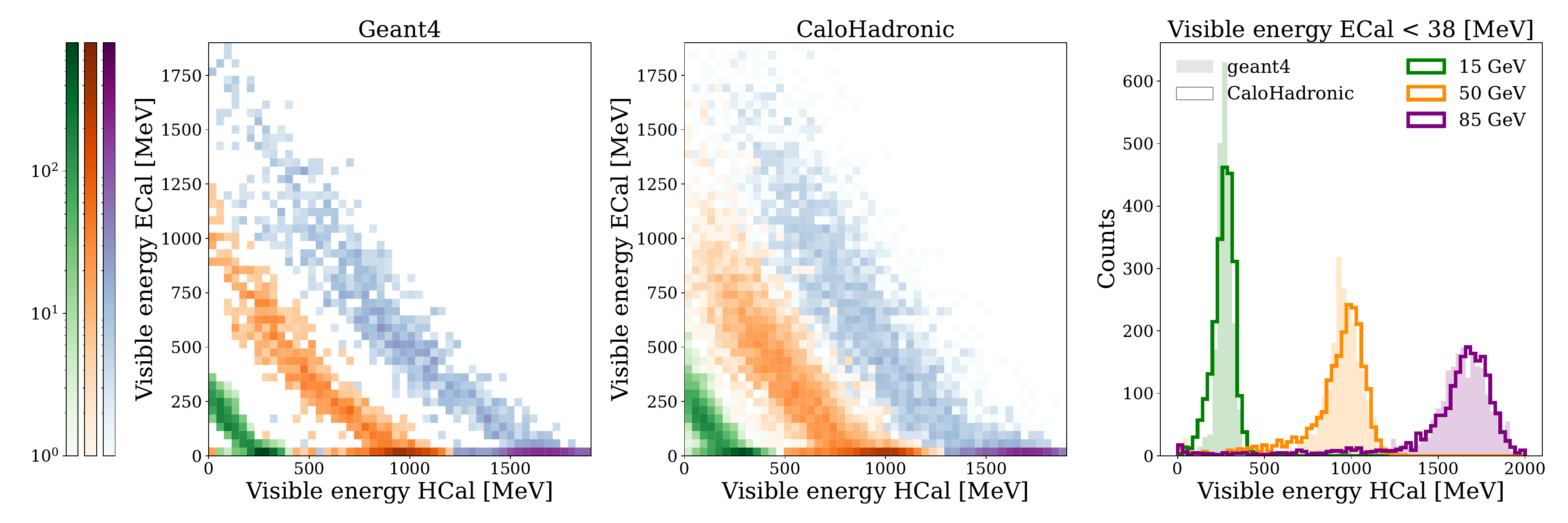}
   \caption{In the two 2D histograms the energy sum in the electromagnetic calorimeter and the energy sum in the hadronic calorimeter is shown for 15~GeV (green), 50~GeV (orange) and 85~GeV (purple) showers for both Geant4 (left) and \cc (center). The color represents the number of showers. Here $5,000$ showers were used. Geant4 showers are sampled within $\pm$ 1 MeV of the target energy; \cc showers are newly generated at the same incident energies.  
   The 1D histogram (on the right) shows the HCal energy sum when the ECal energy sum 
   is below 38~MeV. } 
   \label{fig:corrmultiEn}
\end{figure*}

Figure \ref{fig:corrmultiEn} shows two 2D histograms of the total visible energy deposited in the ECal against the total visible energy deposited in the HCal for incident energies of 15~GeV, 50~GeV and 85~GeV, respectively. The 1D histogram on the right side gives a clearer view of the HCal total visible energy when the ECal total visible energy is below 38~MeV. Most of the 
pion showers lie in the region of lower energy depositions in the ECal and higher energy depositions
in the HCal. This is due to the fact that a large fraction of the pions travel through the ECal before showering in the HCal, where the majority of their energy is deposited. Compared to Geant4, \cc shows a slightly broader band of fluctuations and a slightly different slope in the visible energy distribution of ECal versus HCal. While this could, in principle, affect electron–hadron separation, in practice it does not pose an issue, as high-granularity calorimeters offer several alternative methods to achieve this discrimination.

\subsection{Timing}\label{subsec:timing}

The primary objective of training generative networks on calorimeter 
showers is to achieve a significant speedup compared to Monte Carlo-based 
simulators. To evaluate the performance gain, we conduct a timing 
study to compare the shower generation speed of \cc with that of Geant4. 
The times are computed from three runs, each using incident energies ranging from 10 to 90~GeV in 10~GeV increments, with 100 showers per energy point.

The timing evaluation shown in Table \ref{tab:timing} presents results for both a single AMD EPYC 7402 CPU and an NVIDIA® A100 GPU 80GB. To ensure a fair comparison with Geant4 for the CPU timing, only a single core of the CPU is used.
The mean and standard
deviation of the time per shower over 3 runs for both GPU and
CPU were calculated. The results are presented for various number of function evaluations (NFE) in the generation and for different batch sizes.
Batch sizes of 1 and 16 were chosen because initial studies showed that
in order to generate all the showers, batch sizes of 16, 32 and 64 performed similarly, while a batch size of 128 was slower. This is likely due to the fact that for batch sizes which are too large the GPU cash overflows, resulting in a slowdown.

\begin{table}[ht]
   \small
   \centering
   \resizebox{\textwidth}{!}{ 
   \begin{tabular}{llllcc}
      \toprule
      \textbf{Simulator} &\textbf{Hardware} & \textbf{NFE} & \textbf{Batch Size} & \textbf{Time / Shower [s]} & \textbf{Speed-up} \\
      \midrule 
      Geant4  & CPU           &   & 1   &  2.09 $\pm$  0.05 & $\times$ 1 \\
      \midrule 
      \cc & CPU        & 1 & 1   &   0.591 $\pm$ 0.001 & $\times$ 3.5 \\
           &           &   & 16  &  0.7342 $\pm$ 0.0006 & $\times$ 2.8 \\
           \cline{3-6} 
           &             & 29 & 1   &  17.2 $\pm$  0.1 & - \\ %\times$ 0.12 \\
           &             &    & 16  &  21.37 $\pm$  0.05 & - \\ %$\times$ 0.09 \\
           \cline{3-6} 
           &             & 59 & 1   &  34.8 $\pm$ 0.4 & - \\ %$\times$ 0.06 \\
           &             &    & 16  &  43.3 $\pm$ 0.3 & - \\ %$\times$ 0.05 \\
      \cline{2-6}
           & GPU         & 1 & 1   & 0.0086 $\pm$ 0.0007 & $\times$ 243 \\
           &             &   & 16  & 0.0033 $\pm$ 0.0009 & $\times$ 633  \\
           \cline{3-6}
           &             & 29 & 1   & 0.1978 $\pm$ 0.0007 & $\times$ 11 \\
           &             &    & 16  & 0.0752 $\pm$ 0.0004 & $\times$ 28 \\ 
           \cline{3-6}
           &             & 59 & 1   &  0.3962 $\pm$ 0.0008   & $\times$ 5 \\
           &             &    & 16  & 0.1531 $\pm$ 0.0002 & $\times$ 14 \\ 
       \bottomrule
   \end{tabular}
   }
   \caption{Comparison of the computational performance of \cc to the baseline Geant4 simulator.
   All CPU runs were performed on a single core of an AMD EPYC 7402 CPU, 
   and GPU runs used an NVIDIA® A100 with 80 GB memory. For each configuration the showers are generated
   using incident energies ranging from 10 to 90~GeV in 10~GeV increments, with 100 showers per energy point. The mean and standard deviation is computed across three runs.
   Results show mean $\pm$ std over the 3 runs.} 
   \label{tab:timing}
\end{table}

In table \ref{tab:timing} it can be seen that \cc is faster 
than Geant4 on CPU, and is significantly faster on GPU, for a single forward evaluation. As soon as the number of function evaluations (NFE) increases, i.e. 29 or 59, the model is slower than Geant4 on CPU while on GPU it is still faster.
Following \cite{karras2022edm}, each Heun step requires two function evaluations, except for the final iteration, where a first-order Euler step is used instead of the second-order Heun method, resulting in only one function evaluation at the end.
A second study of how each 
of the three components of \cc contributes
to the overall generation time was performed. Around $80\%$ is due to \hcalD, 
around $18\%$ to \ecalD and around $2\%$ to \pointsFM. 
This is due to the fact that \hcalD and \ecalD are both based on transformers 
which are known to be computationally expensive. Here the advantages of a smaller model for \ecalD with respect to \hcalD are clear. 

For both \ecalD and \hcalD models, sampling was performed using 59 NFE. As a result, generating each event requires multiple forward evaluations of the model, which directly impacts inference speed. This computational cost highlights a key avenue for future improvement—namely, the reduction of the NFE through the use of accelerated samplers such as consistency models or distillation techniques.

\section{Reconstruction with Pandora Particle Flow}\label{section:reco}

For a fast calorimeter shower model to be used to produce simulated data for physics analyses, it must ultimately be interfaced with the reconstruction chain used by a given experiment. Detectors designed for operation at future $e^{+}e^{-}$ colliders are typically optimized for the particle flow approach to reconstruction. The goal of this approach is to individually reconstruct each particle present in an event. To this end, measurements of charged particles are made in the tracking system, while measurements of neutral particles are made in the calorimeters. This drives the need for high granularity in the calorimeter systems, which is essential to correctly separate clusters of hits associated with neutral particles from those associated with charged particles.

In this study, we employ the state-of-the-art \Pandora particle flow algorithm \cite{Thomson:2009rp, Marshall:2015rfa}, which is used in the ILD standard reconstruction chain. The simulated hits produced by either Geant4 or the generative model first undergo a digitization procedure, which takes into account detector effects such as noise from the electronics. A two-step calibration procedure is then applied to the hits, after which the sum off all the hits in the shower corresponds to the energy of the incident particle. It is these digitized/calibrated calorimeter hits, together with tracks formed in the tracking system which are provided as input to \Pandora.

\Pandora then applies a complex series of pattern recognition algorithms, which ultimately aim to correctly form calorimeter clusters and assign corresponding tracks where appropriate. The output produced by \Pandora is a list of reconstructed objects referred to as \textit{Particle Flow Objects} (PFOs), each of which contains information about the particle's four-momentum and ID.

\subsection{Methods}\label{section:reco:Methods}

In order to be able to apply the standard reconstruction suite of ILD to showers produced by the model, it is necessary to be able to correctly place the energy deposits into the detector geometry. We therefore use the \textsc{DDML} library \cite{McKeown:2023QR, McKeown:2024jeq} to combine the model output with a full simulation application. We fire a $\pi^{+}$ with an energy of $50$~GeV from a particle gun positioned at ($-5, 0, -15$)~mm\footnote{This position is chosen to avoid the TPC cathode, which is positioned at $z=0$ in the global ILD coordinate system} in the global ILD coordinate system, with a direction chosen to produce a perpendicular incidence at the calorimeter face. This takes into account the slight curvature of the track. The particles are not produced directly at the face of the calorimeter, as it is necessary for a track to be present in the tracking system for \Pandora to apply the correct sequence of algorithms for $\pi^{+}$ reconstruction. However, the energy of the pion used for conditioning of the model is extracted directly at the face of the calorimeter, to match the training scenario. 

\begin{figure*}[htbp]
   \centering 
   \includegraphics[width=75mm]{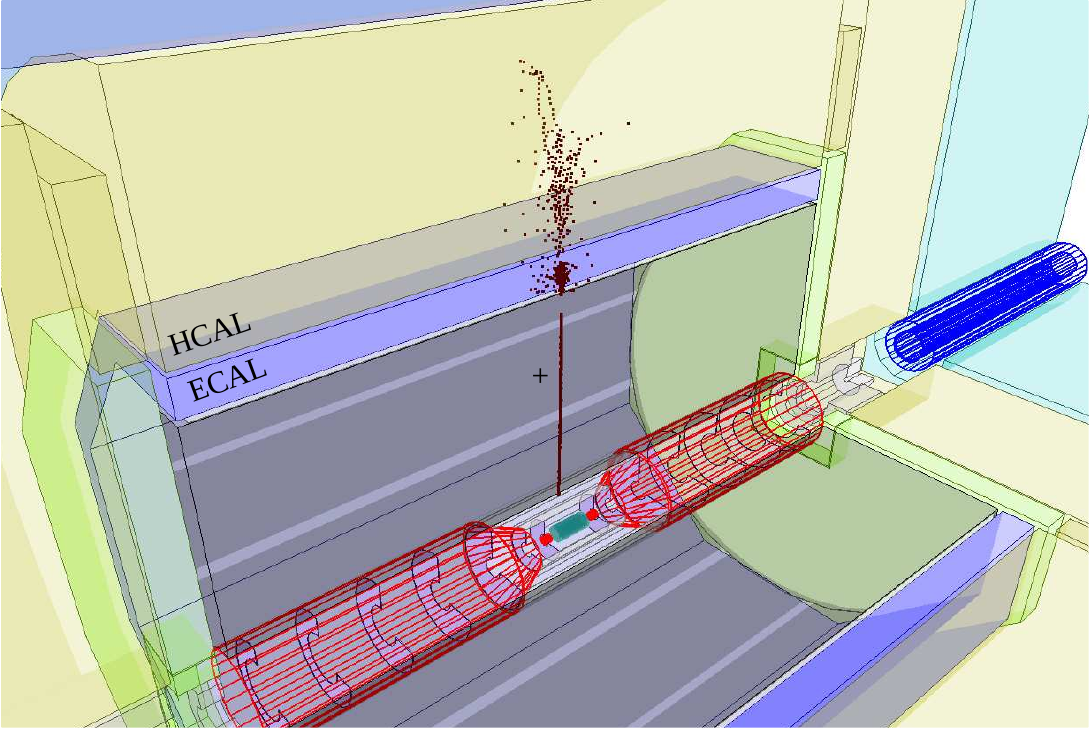}
   \includegraphics[width=75mm]{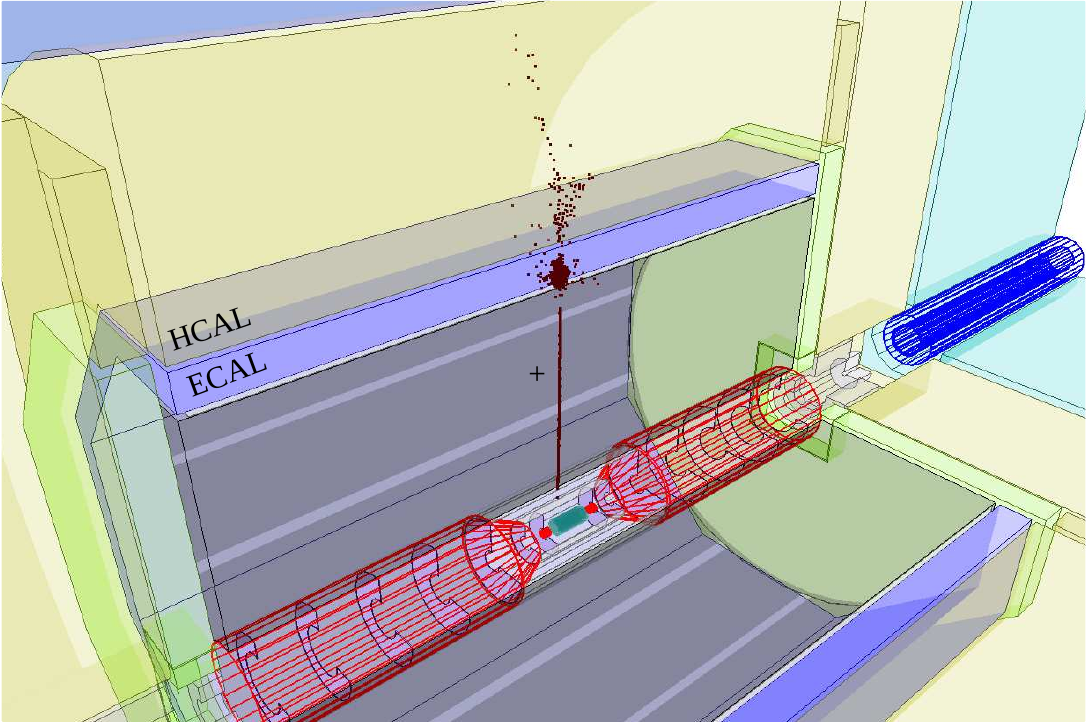}
   \caption{Event displays of Pandora PFOs reconstructed from a Geant4 shower (left), and a \cc shower (right).}
   \label{fig:Pandora_PFOs}
\end{figure*}

Figure \ref{fig:Pandora_PFOs} shows example event displays of the Pandora PFOs reconstructed for a Geant4 shower (left) and for a \cc shower (right). In total, $2000$ such events are generated for both Geant4 and the model. After cuts were applied on the quality of the reconstructed track, a total of 1929 events for \cc and 1941 events for Geant4 remain.

\subsection{Results}\label{section:reco:Results}

\begin{figure*}[htbp]
   \centering 
   \includegraphics[width=150mm]{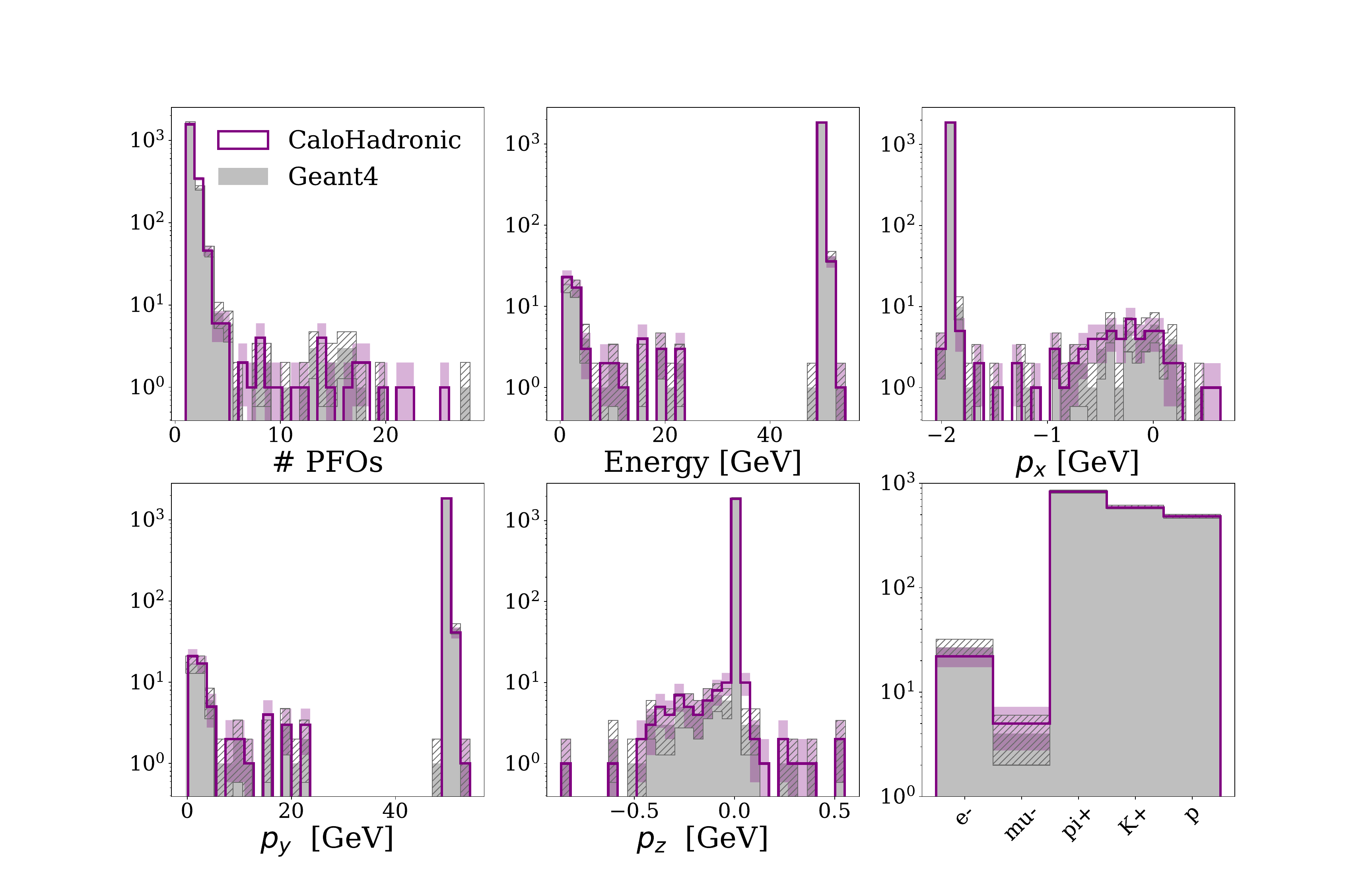}
   \caption{PFO-level observables after reconstruction of pions with \Pandora for Geant4 (gray) and \cc (purple). Observables include distributions for the number of PFOs reconstructed per event (\# PFOs, top left), PFO energy (top middle), PFO momentum in $x$ ($p_x$, top left), PFO momentum in $y$ ($p_y$, bottom left), PFO momentum in $z$ ($p_z$, bottom middle), and the type of reconstructed particle (bottom right).}
   \label{fig:reco}
\end{figure*}

We now analyse the effects of reconstruction by studying the performance of the model compared to Geant4 in terms of key physics observables after reconstruction. The observables studied are the energy of the PFO, its momentum in each direction ($p_x$, $p_y$, $p_z$) and the reconstructed particle type. Figure \ref{fig:reco} shows distributions for each of these observables for all PFOs reconstructed across all events, as well as the number of PFOs (\# PFOs) reconstructed in each event. The majority of PFOs are correctly reconstructed as $\pi^{+}$, however, a significant fraction (~57$\%$) are incorrectly reconstructed as other species of charged hadrons, namely $K^{+}$ and $p$. A small fraction are incorrectly reconstructed as $e^{-}$ and $\mu^{-}$. However, for all observables \cc provides an impressive modeling of the various observables within uncertainties, including reproducing the deficiencies present in the reconstruction that are also present when simulating with Geant4.

\begin{figure*}[htbp]
   \centering 
   \includegraphics[width=150mm]{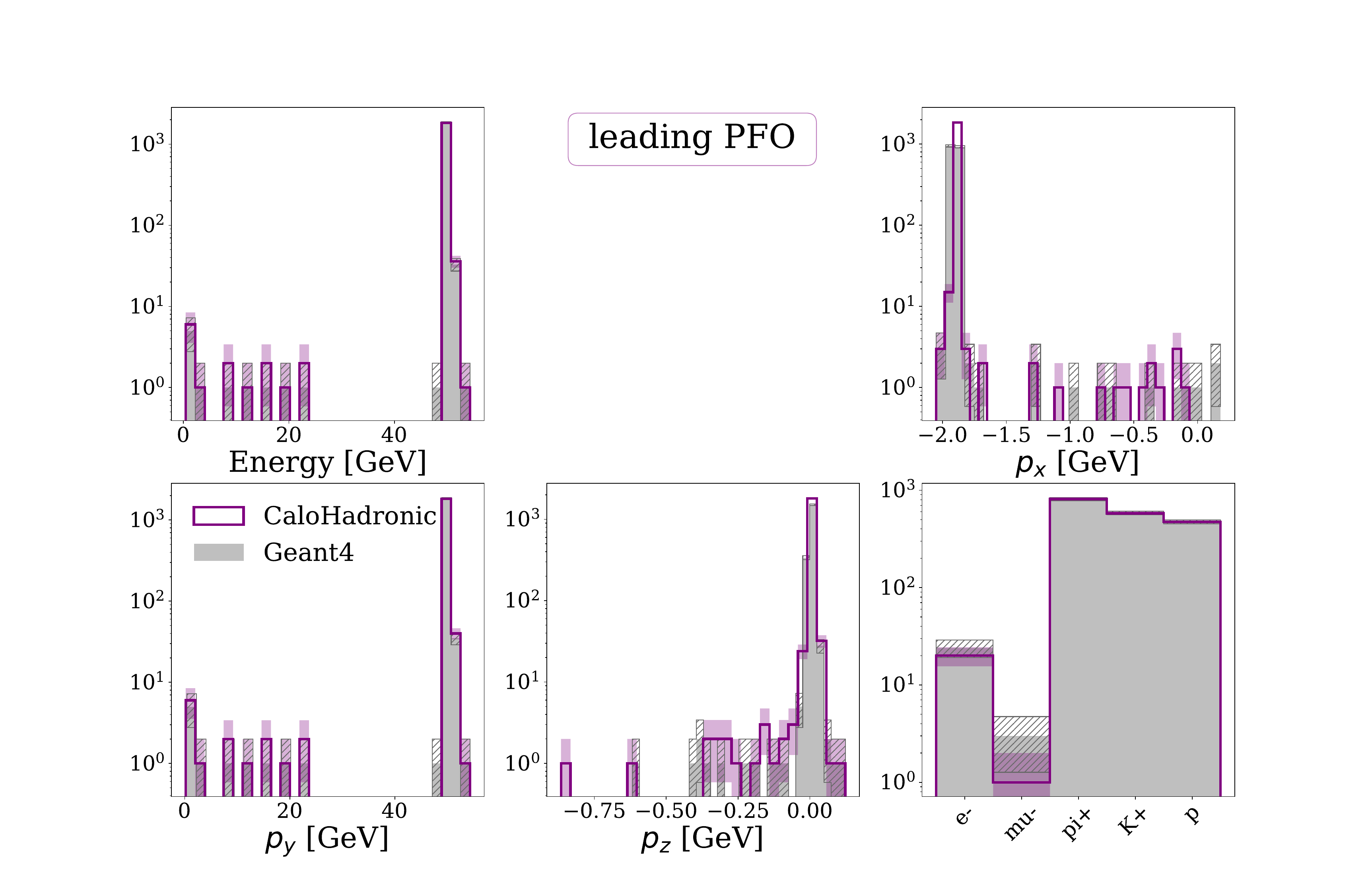}
   \caption{PFO-level observables after reconstruction of pions with \Pandora for Geant4 (gray) and \cc (purple). Only 
   the leading PFO per event is shown. Observables include distributions for the leading PFO energy (top middle), leading PFO momentum in $x$ ($p_x$, top left), leading PFO momentum in $y$ ($p_y$, bottom left), leading PFO momentum in $z$ ($p_z$, bottom middle), and the type of the leading reconstructed particle (bottom right).}
   \label{fig:reco1}
\end{figure*}

Due to the complexity of hadronic showers, it is possible for multiple PFOs to be reconstructed from a single event. For this reason, we also show the reconstruction observables for the leading (i.e. highest energy) PFO per event in Figure \ref{fig:reco1}. \cc again provides an impressive modeling of post-reconstruction PFO level observables.

\section{Discussion}
Overall, \cc reproduces the main features of hadronic showers with very good agreement to Geant4. The longitudinal and radial profiles, as well as the center of gravity distributions, all show that the model captures the essential shower development. Small deviations are visible, but they remain within acceptable limits for physics applications as demonstrated by the excellent post-reconstruction performance at the single shower level.

In the longitudinal profile, the alternating structure in the ECal layers is well described by Geant4 but slightly less so by \cc. While this effect is relevant for the reconstruction of electromagnetic showers, for which there are multiple other generative models, it only has a negligible effect on reconstruction for hadronic showers. 

In the radial direction, \cc tends to generate showers that are slightly narrower than those from Geant4. This could, in principle, reduce the separation power of particle flow reconstruction where showers overlap. 

Looking at the center of gravity (CoG), the model describes the distributions in all three directions ($x$, $z$, $y$) well. Minor differences are observed: the transverse CoGs ($x$, $z$) are slightly narrower, while along the shower axis ($y$) the distribution is shifted by a small amount. 

As noted in the cell energy spectrum (figure \ref{fig:dist1}), \cc slightly overestimates the hit energy, raising the concern that this might distort the total energy distribution. The comparisons presented later in figure \ref{fig:hits} show that the impact of this effect is very subtle: while present, it does not significantly alter the global energy of the showers.

The number of hits and total visible energy per shower are also modeled with high fidelity across a wide range of incident energies. Both the resolution and linearity are preserved, and the comparison at fixed energies (15, 50, 85 GeV) confirms that \cc follows the Geant4 reference closely.

Correlation studies provide a more stringent test. The joint distributions of observables, such as energy sum versus center of gravity or number of hits versus energy sum, show overall agreement. The difference plots reveal systematic trends: \cc sometimes overpopulates the central regions while underpopulating the tails, and at fixed numbers of hits it slightly underestimates the energy sum compared to Geant4. Still, these deviations are modest, and the main correlations are well preserved.

The comparison of visible energy deposited in the ECal and HCal demonstrates that \cc captures the characteristic pion behavior. The model reproduces the global slope and fluctuations, with only slightly larger variations than Geant4. Importantly, while such differences could affect electron–hadron separation in principle, modern highly granular calorimeters provide several independent handles to achieve this task, so in practice it is not a limiting factor.

After downstream processing with PandoraPFA, \cc reproduces key reconstructed observables—including PFO energy, momentum, multiplicities, and particle ID—with accuracy comparable to Geant4. Both models show similar reconstruction inefficiencies, and at the event level \cc closely matches Geant4 for all PFOs as well as the leading PFO. This confirms that small deviations seen at the hit and shower level do not propagate into significant mismodeling after reconstruction at the single particle level. Future work could study the performance of \cc in multi-particle environments, where overlapping showers would provide more stringent tests of model performance. Moreover, incorporating the time dimension would be particularly relevant for studying hadronic shower development and pile-up effects. Ultimately, the model should be tested at the level of observables for higher level reconstructed objects (e.g. jets) in full physics events.

On CPU, \cc achieves a modest speed-up over Geant4, while on GPU the gain reaches up to three orders of magnitude. Higher numbers of function evaluations (NFE) reduce this advantage, but even with 59 NFE the GPU remains faster. Most of the runtime comes from the transformer-based diffusion modules, suggesting that accelerated sampling techniques could bring further improvements.

Taken together, these results show that \cc is able to reproduce both single-variable distributions and correlations with good accuracy. The small deviations observed are well understood and do not compromise the usefulness of the model for realistic physics studies. Importantly, \cc generates visually faithful shower shapes, particularly sub-shower tracks, which are crucial for Pandora Particle Flow to correctly reconstruct hadronic showers.

\section{Conclusion}

To turn generative models from proof-of-concept studies into useful surrogates to be used in production, they need to be able to handle the various complexities of realistic calorimeters.
This work demonstrates for the first time how a point-cloud based generative model can be extended to simultaneously simulate the ECal and HCal parts of a hadronic shower, emphasizing the flexibility of this approach.
Both individual shower properties, as well as correlations (also across ECal and HCal components), are described well. A classifier test shows that --- as for the generation of partial showers --- the generative fidelity needs to be further improved. This is especially visible in the fine grained track-like structures observed in the HCal. 
However the degree of agreement between generative output and ground-truth processing with reconstruction algorithms is of a similar or slightly better quality however the degree of agreement between generative output and ground-truth is of a similar or slightly better quality when also considering down-stream processing with reconstruction algorithms.

Finally, the demonstrated strategy of sequentially conditioning several generative models might also be applied to other complex generative simulation tasks beyond calorimeters.

\section{Acknowledgments}
We thank William Korcari for early comments to this work. 
We want to thank Freya Blekman for valuable comments on the manuscript.
This research was supported in
part by the Maxwell computational resources operated at Deutsches Elektronen-Synchrotron DESY,
Hamburg, Germany. This project has received funding from the European Union’s Horizon 2020
Research and Innovation program under Grant Agreement No 101004761. We acknowledge
support by the Deutsche Forschungsgemeinschaft under Germany’s Excellence Strategy – EXC
2121 Quantum Universe – 390833306 and via the KISS consortium (05D23GU4, 13D22CH5)
funded by the German Federal Ministry of Research, Technology and Space (BMFTR) in the ErUM-Data action plan. P.M. has benefited from support by the CERN Strategic R$\&$D Programme on Technologies for Future Experiments \cite{EPRD}. 

\newpage
\bibliographystyle{JHEP}
% \section{Bibliography}
\addcontentsline{toc}{section}{Bibliography}
\bibliography{bibliography.bib}

\newpage
\appendix
\section{Model details and
hyper-parameters}\label{sec:appA}

In tables \ref{tab:training_config} and \ref{tab:pointsFM_config} a detailed view of the hyper-parameters used in the trainings of \pointsFM \ecalD and \hcalD is presented.

\begin{table}[htbp]
    \centering
    \resizebox{\textwidth}{!}{ 
    \begin{tabular}{llll}
    \toprule
    Category & \textbf{Hyperparameter} & \textbf{\ecalD} & \textbf{\hcalD} \\
    \midrule
    \multirow{6}{*}{\textsc{\large{General}}} 
    & \textbf{only\_HCal} & True & False \\
    & \textbf{only\_ECal} & False & True  \\
    & \textbf{Seed} & 45 & 455 \\
    & \textbf{Training Hardware} & 1 x NVIDIA® A100 & 4 x NVIDIA® A100 \\
    & \textbf{Optimizer} & AdamMini & AdamMini \\
    & \textbf{Loss Type} & EDM-Monotonic & EDM-Monotonic\\
    \midrule
    \multirow{4}{*}{\textsc{\large{EDM features}}} 
    & \textbf{KL Weight} & $10^{-3}$ & $10^{-3}$ \\
    & \textbf{KLD Min} & 1.0 & 1.0 \\
    & \textbf{Schedule Mode} & Quadratic & Quadratic \\
    &\textbf{EMA Type} & Inverse, Power=0.6667, Max=0.9999 & Inverse, Power=0.6667, Max=0.9999 \\
    \midrule
    \multirow{6}{*}{\textsc{\large{Transformer}}} 
    & \textbf{Transformer Decoder Layers} & 0 & 2 \\
    & \textbf{Transformer Encoder Layers} & 3 & 4 \\
    & \textbf{Embedding Dimension} & 128 & 128 \\
    & \textbf{Number of heads} & 8 & 16 \\
    & \textbf{Feedforward dim} & 512 & 512 \\
    & \textbf{Dropout Rate} & 0.1 & 0.1 \\
    \midrule
    \multirow{5}{*}{\textsc{\large{Fuorier Layer}}}
    & \textbf{Include Input} & True & True \\
    & \textbf{Max Frequency} & 16 & 16 \\
    & \textbf{Number of Frequencies } & 32 & 32\\
    & \textbf{Log Sampling} & True & True \\
    & \textbf{periodic Functions} & $[sin, cos]$ & $[sin, cos]$\\
    \midrule
    \multirow{8}{*}{\textsc{\large{Data}}} 
    & \textbf{Granularity} & x9 & x9 \\
    &\textbf{Features} & 4 & 4 \\
    &\textbf{ECal Features} & - & 4 \\
    &\textbf{Cond. Features}  & $30+1$ & $48+1$ \\
    &\textbf{Cond. Normalization} & True & True \\
    &\textbf{Log energy} & True & True \\
    &\textbf{Batch Size} & 32 & 32 \\
    &\textbf{Dataloader Workers} & 40 & 40\\
    \midrule
    \multirow{4}{*}{\textsc{\large{Scheduler}}}
    &\textbf{LR Scheduler} & OneCycleLR & OneCycleLR\\
    &\textbf{LR Start / Max / End} & $3\cdot10^{-5}$ / $3\cdot10^{-4}$ / $1\cdot10^{-6}$ & $3\cdot10^{-5}$ / $3\cdot10^{-4}$ / $1\cdot10^{-6}$\\
    &\textbf{Warmup Steps} & 300k & 500k  \\
    &\textbf{Total number of gradient steps} & 2M & 3M  \\
    \midrule
    \multirow{6}{*}{\textsc{\large{Sampling}}}
    &\textbf{Sigma Data} & 0.5 & 0.5 \\
    & \textbf{Sigma Sampling} & Lognormal($\mu$=0, $\sigma$=1) & Lognormal($\mu$=0, $\sigma$=1)\\
    &\textbf{ODE Solver} & Heun & Heun\\
    &\textbf{\# Sampling Steps} & 30 & 30 \\
    &\textbf{Sigma Min / Max} & 0.01 / 10.0 & 0.01 / 10.0 \\
    &\textbf{Rho / s\_churn / s\_noise} & 7.0 / 0.0 / 1.0 & 7.0 / 0.0 / 1.0 \\
    \bottomrule
    \end{tabular}
    }
    \caption{Summary of \ecalD and \hcalD Training Configuration}
    \label{tab:training_config}
    \end{table}

    \begin{table}[htbp]
        \centering
        \resizebox{0.55\textwidth}{!}{ 
        \begin{tabular}{lll}
        \toprule
        \textbf{Category} & \textbf{Hyperparameter} & \textbf{\pointsFM} \\
        \midrule
        \multirow{4}{*}{\textsc{\large{Data}}}
        & \textbf{Train/Val Split} & 90\% / 10\% \\
        & \textbf{Pin Memory} & True \\
        & \textbf{Workers} & 20 \\
        & \textbf{Shuffle} & True \\
        \midrule
        \multirow{4}{*}{\textsc{\large{Architecture}}}
        & \textbf{Num Inputs} & 78 \\
        & \textbf{Conditioning Inputs} & 1 \\
        & \textbf{Time Embedding Dim} & 6 \\
        & \textbf{Hidden Dims} & [128, 256, 512, 256, 128] \\
        \midrule
        \multirow{7}{*}{\textsc{\large{Training}}}
        & \textbf{Device} & NVIDIA® A100 \\
        % & \textbf{Seed} & 36 \\
        & \textbf{Optimizer} & Adam-mini \\
        & \textbf{Scheduler} & OneCycleLR \\
        & \textbf{Max Learning Rate} & 0.001 \\
        & \textbf{Warm up epochs} & 900 \\
        & \textbf{Batch Size} & 3000 \\
        & \textbf{Epochs} & 3000 \\
        \midrule
        \multirow{2}{*}{\textsc{\large{Generation}}}
        & \textbf{ODE Solver} & Heun \\
        & \textbf{\# Sampling Steps} & 200 \\
        \bottomrule
        \end{tabular}
        }
        \caption{Configuration summary for the \pointsFM.}
        \label{tab:pointsFM_config}
    \end{table}

\section{Compression block}\label{sec:appSCL}
Taking inspiration from \cite{ji2024latent}, a compression block was implemented. The compression block is a simple neural network block designed to reduce a variable-sized set of point features into a fixed number of representative ``tokens", 10 in this case. 
The process happens in two main steps:
\begin{itemize}
    \item First, the model looks at each input point and decides how much it should contribute to each token. It does this firstly by applying a linear layer to end up with the token's dimensionality and then using a softmax function. The outputs are per-point weights that tell the model how important each point is for each token with a probability value.
    \item A linear layer is applied to the input to map it to the output dimensionality. In this case the output dimension is the same as the input one, i.e. the number of features (4).  
    \item After that, the compression block maps the points to the tokens. It does this by computing a weighted average of each point using the weights from step 1 (how important each point was for that token).  
\end{itemize}
The result is a fixed-size output (one vector per token) that summarizes the input set.

\section{\texorpdfstring{3D $\pi^{+}$ showers}{3D pi showers}}\label{sec:appB}

\begin{figure*}
    \centering 
    \includegraphics[width=55mm]{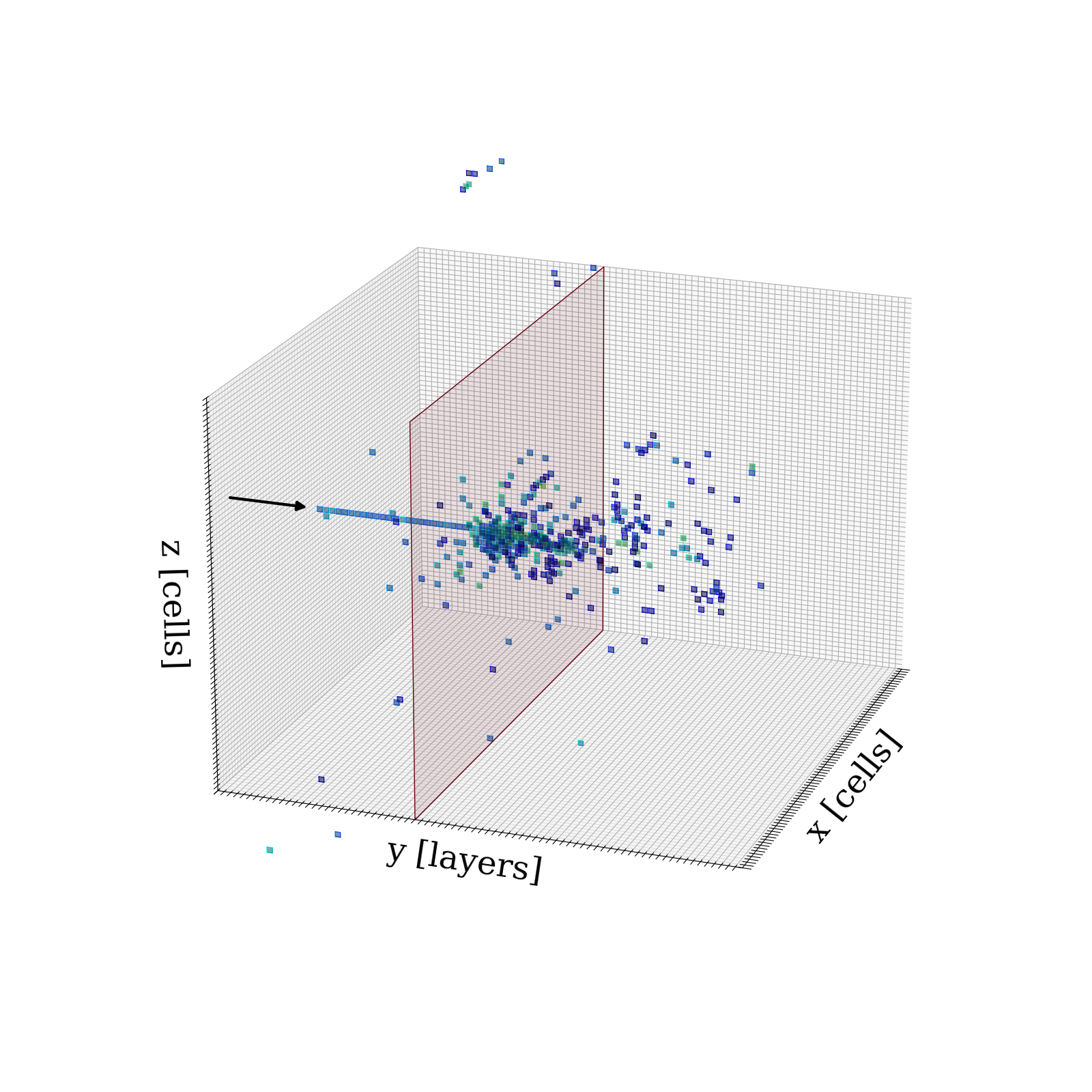}
    \hspace{-11mm}
    \includegraphics[width=55mm]{figures/appendix/3d_50/3d_shower_real.png}
    \hspace{-11mm}
    \includegraphics[width=55mm]{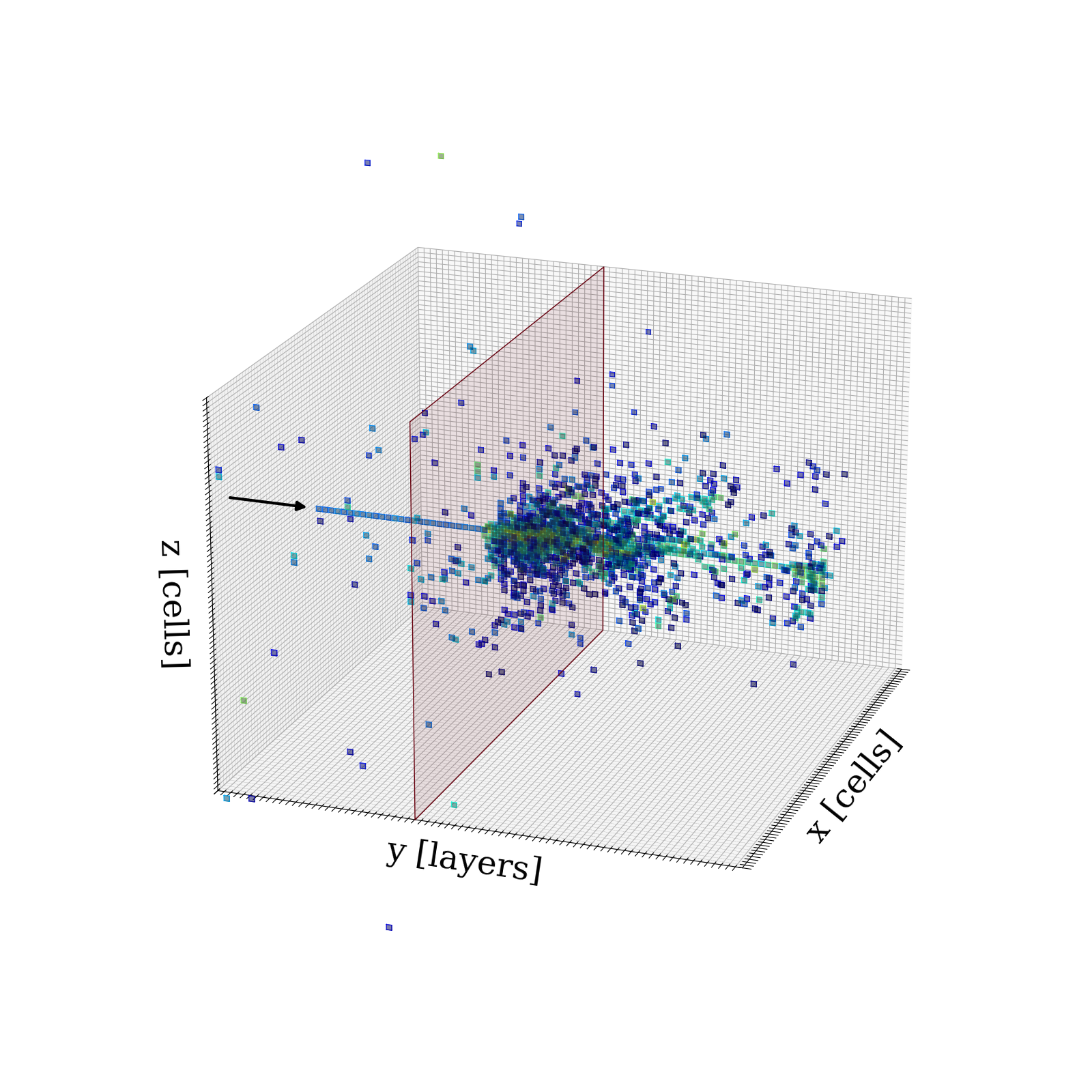}
    \caption{3D view of a 15~GeV (left), a 50~GeV (center), and a 85~GeV (right) Geant4 shower. The color represents the energy deposition in the cells. The red plane represents the division between ECal and HCal.}
    \label{fig:3d_g4}
 \end{figure*}

In this section, we present additional 3D plots of $\pi^{+}$ showers to illustrate the state of the shower-structure simulation. Figure \ref{fig:3d_g4} shows examples of showers simulated by Geant4 with incident energies of 15~GeV, 50~GeV and 85~GeV.

\begin{figure*}
    \centering 
    \includegraphics[width=55mm]{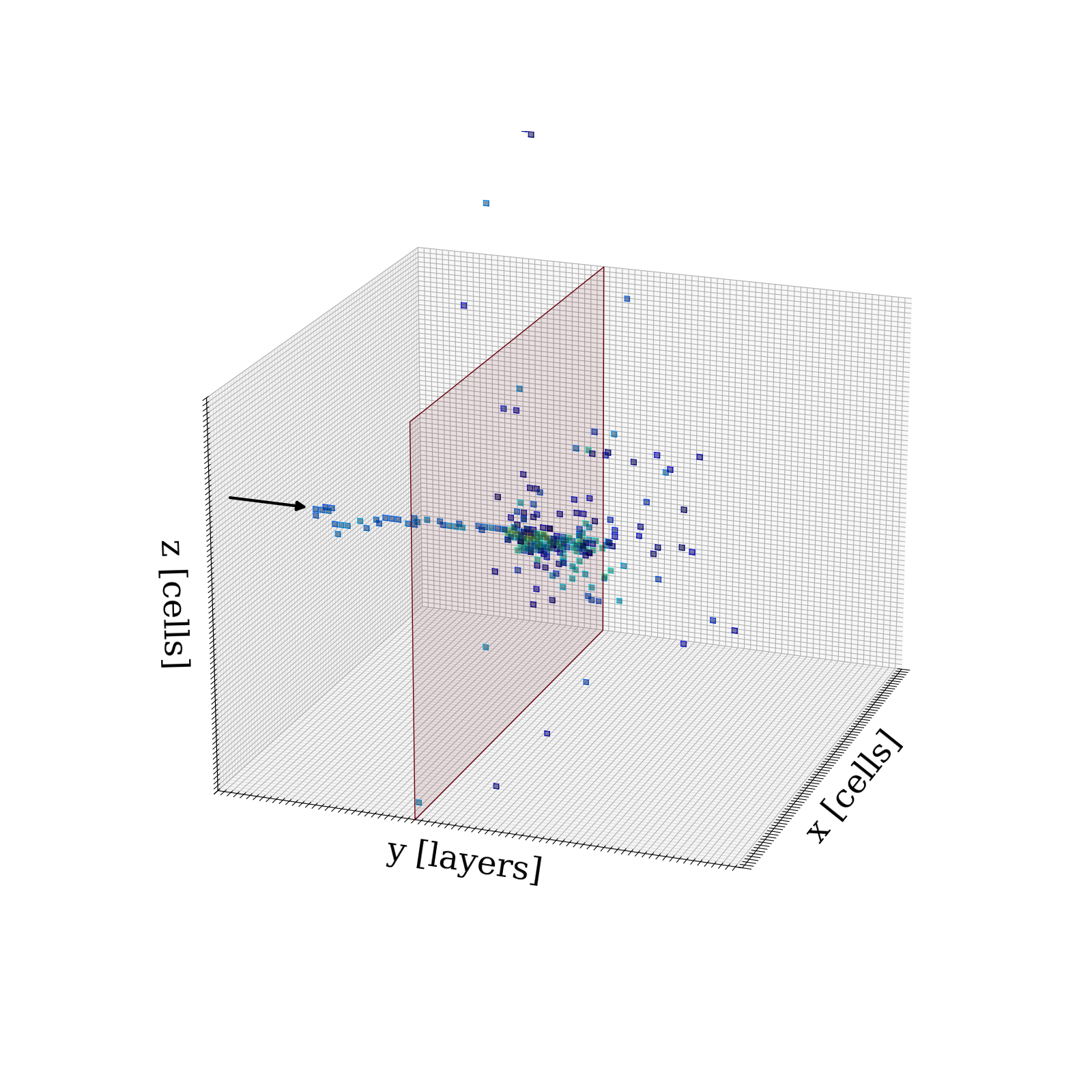}
    \hspace{-11mm}
    \includegraphics[width=55mm]{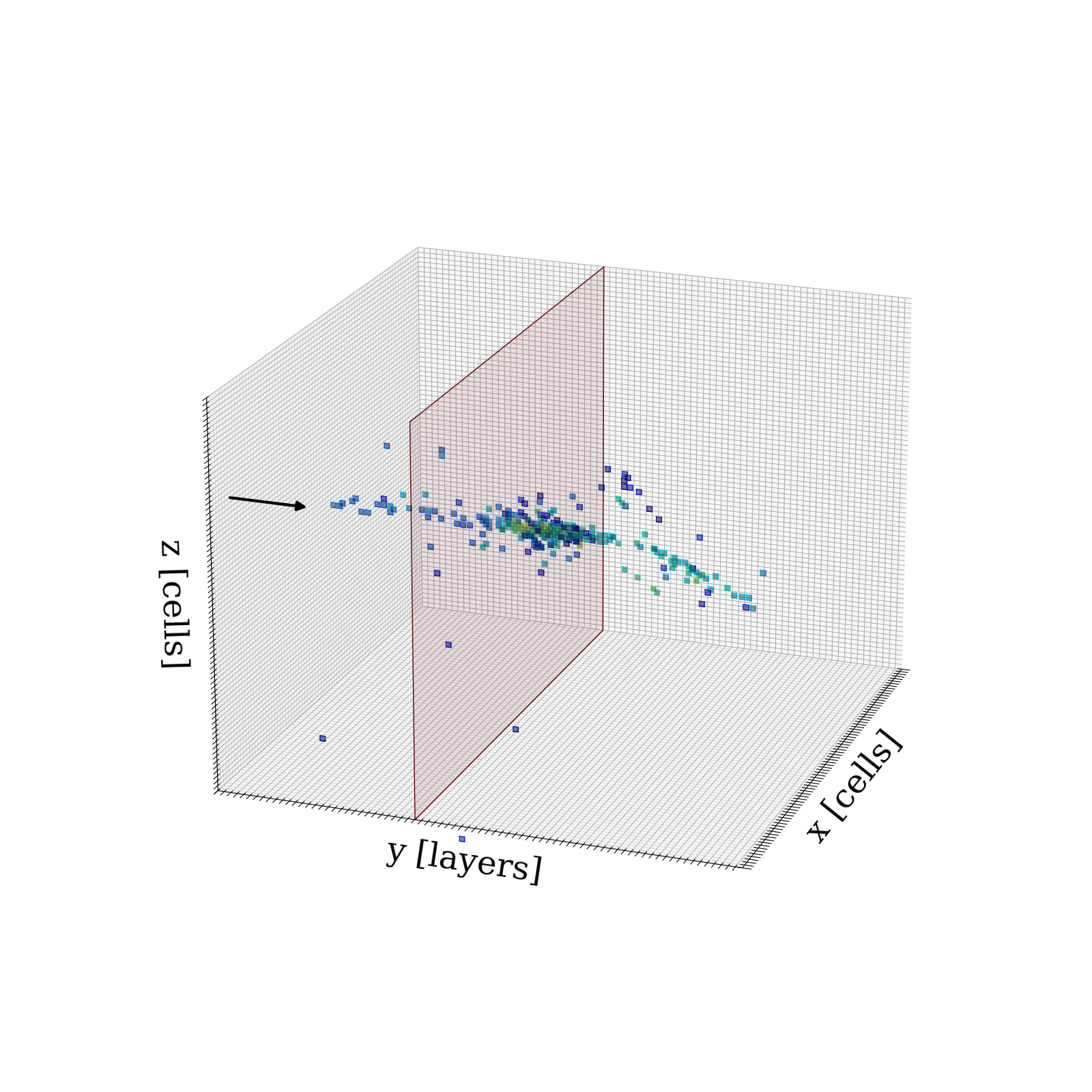}
    \hspace{-11mm}
    \includegraphics[width=55mm]{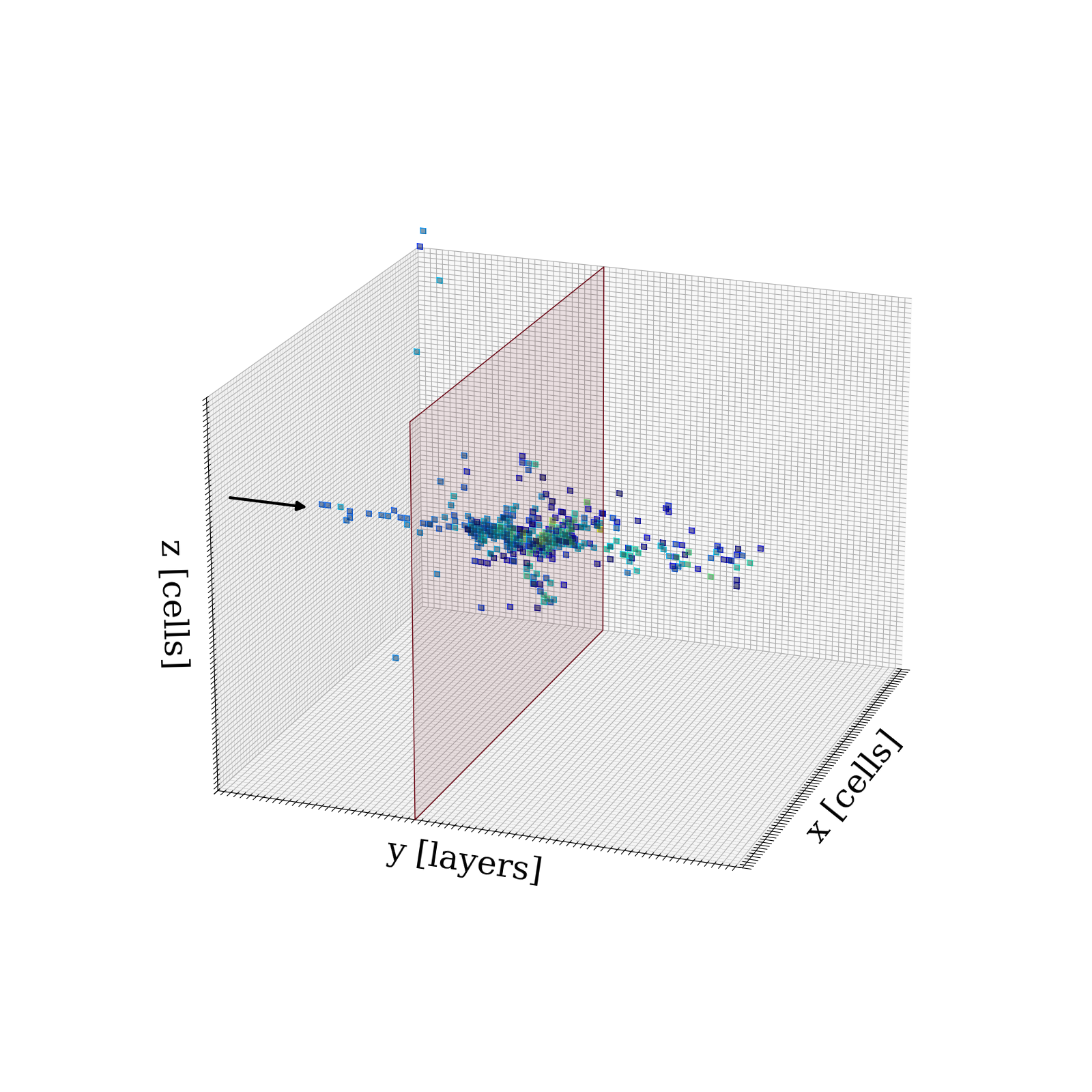}
    \caption{3D view of three 15~GeV \cc showers.
    The color represents the energy deposition in the cells. The red plane represents the division between ECal and HCal.}
    \label{fig:3d_15}
 \end{figure*}

\begin{figure*}
    \centering 
    \includegraphics[width=55mm]{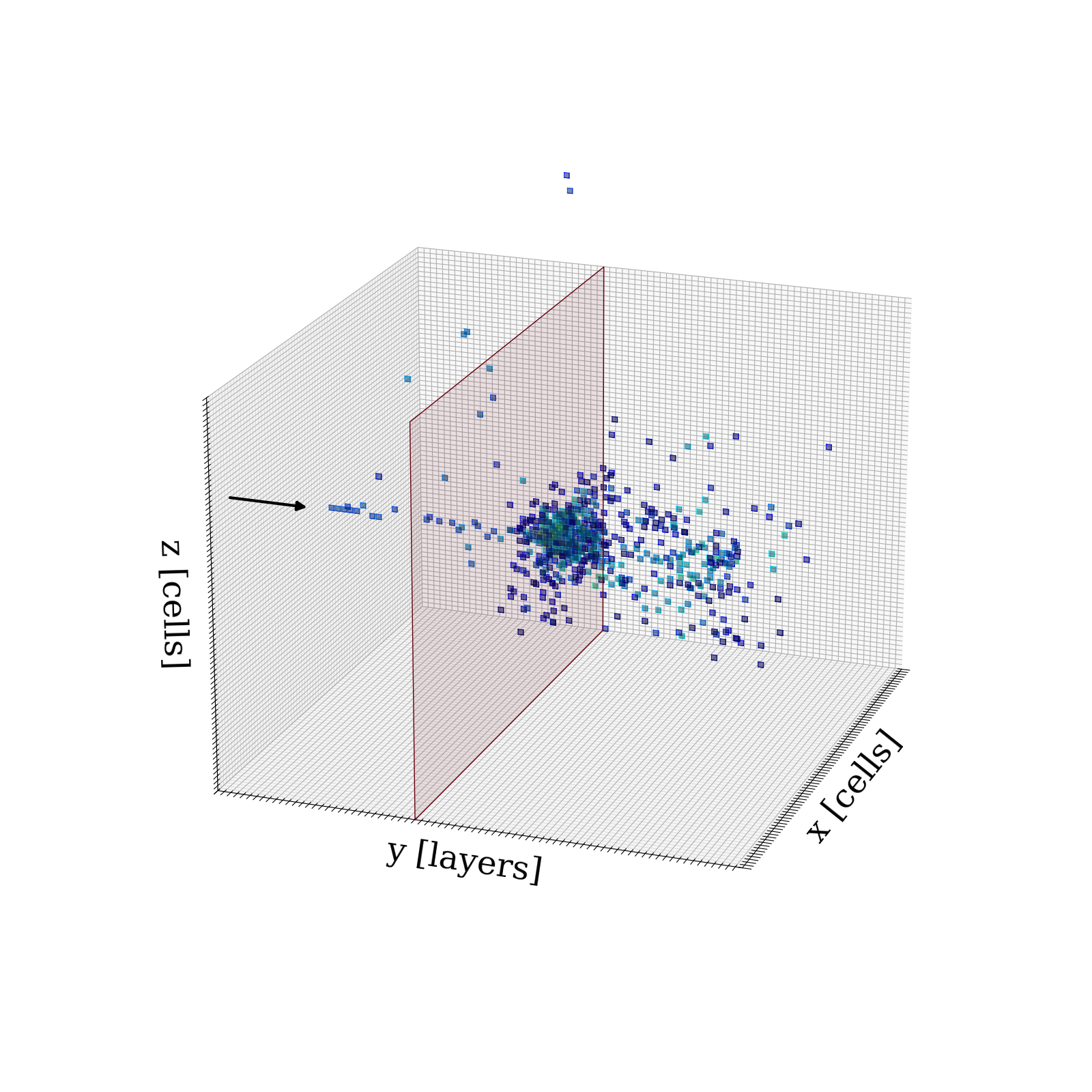}
    \hspace{-11mm}
    \includegraphics[width=55mm]{figures/appendix/3d_50/3d_shower_fake_2503.png}
    \hspace{-11mm}
    \includegraphics[width=55mm]{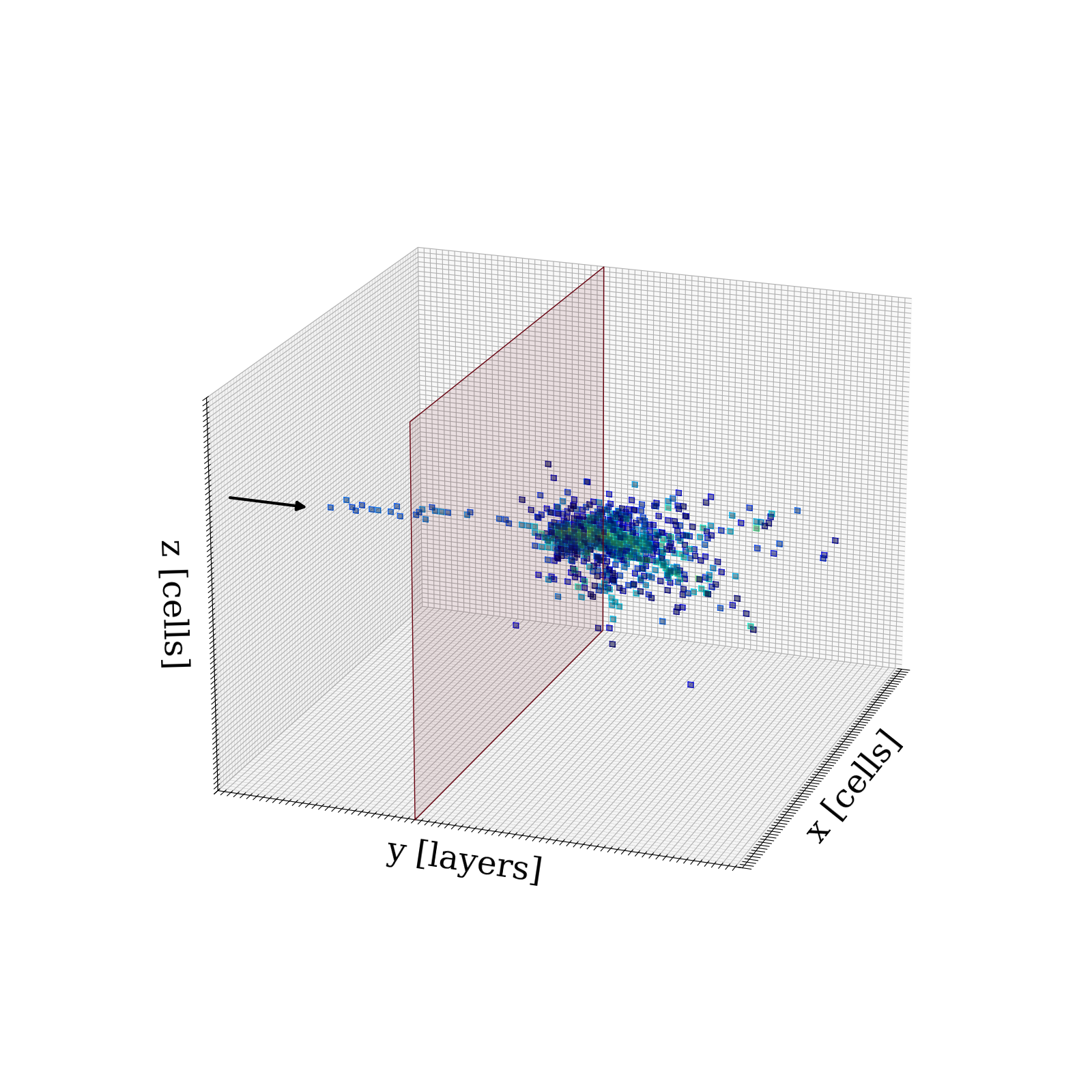}
    \caption{3D view of three 50~GeV \cc showers.
    The color represents the energy deposition in the cells. The red plane represents the division between ECal and HCal.}
    \label{fig:3d_50}
 \end{figure*}

\begin{figure*}
    \centering 
    \includegraphics[width=55mm]{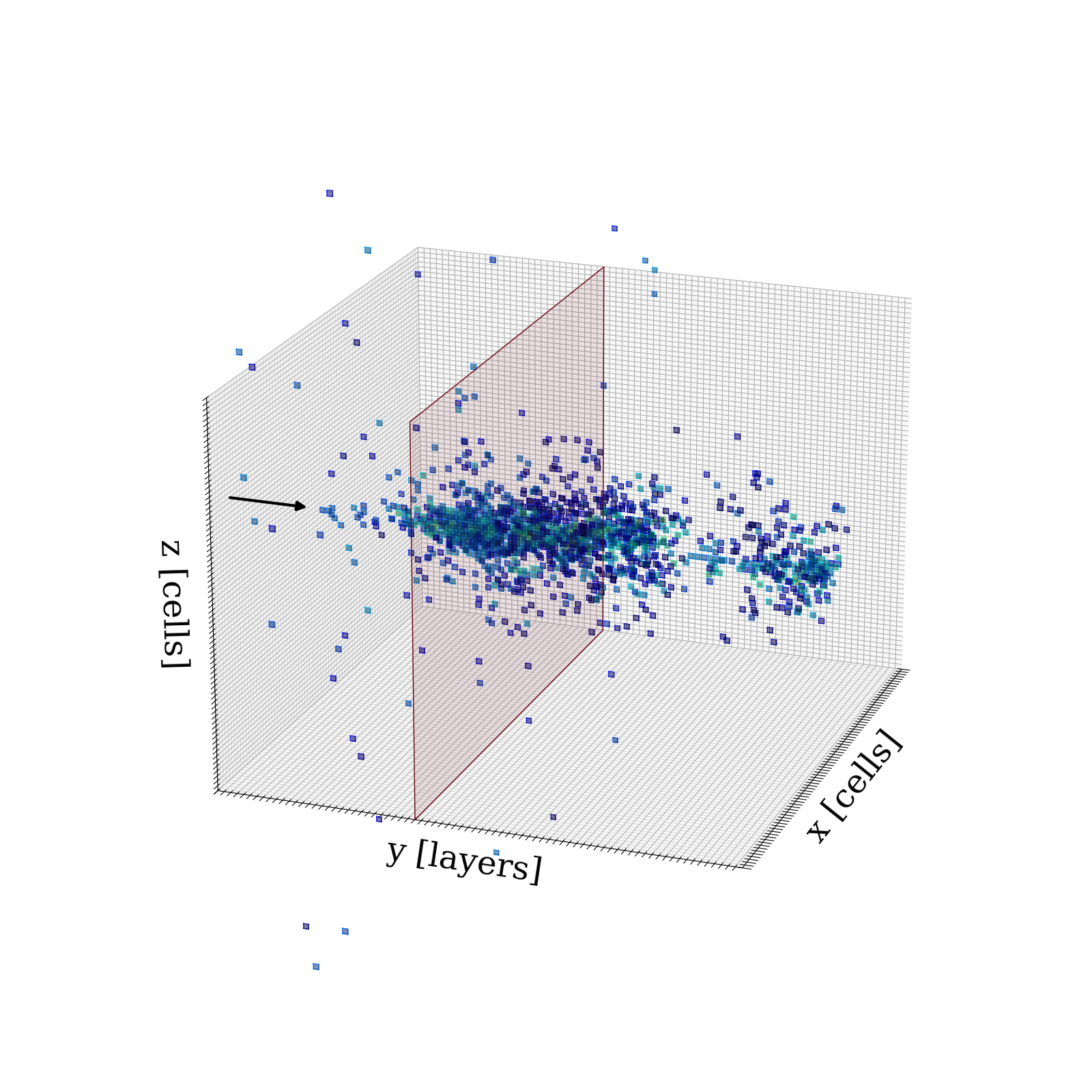}
    \hspace{-11mm}
    \includegraphics[width=55mm]{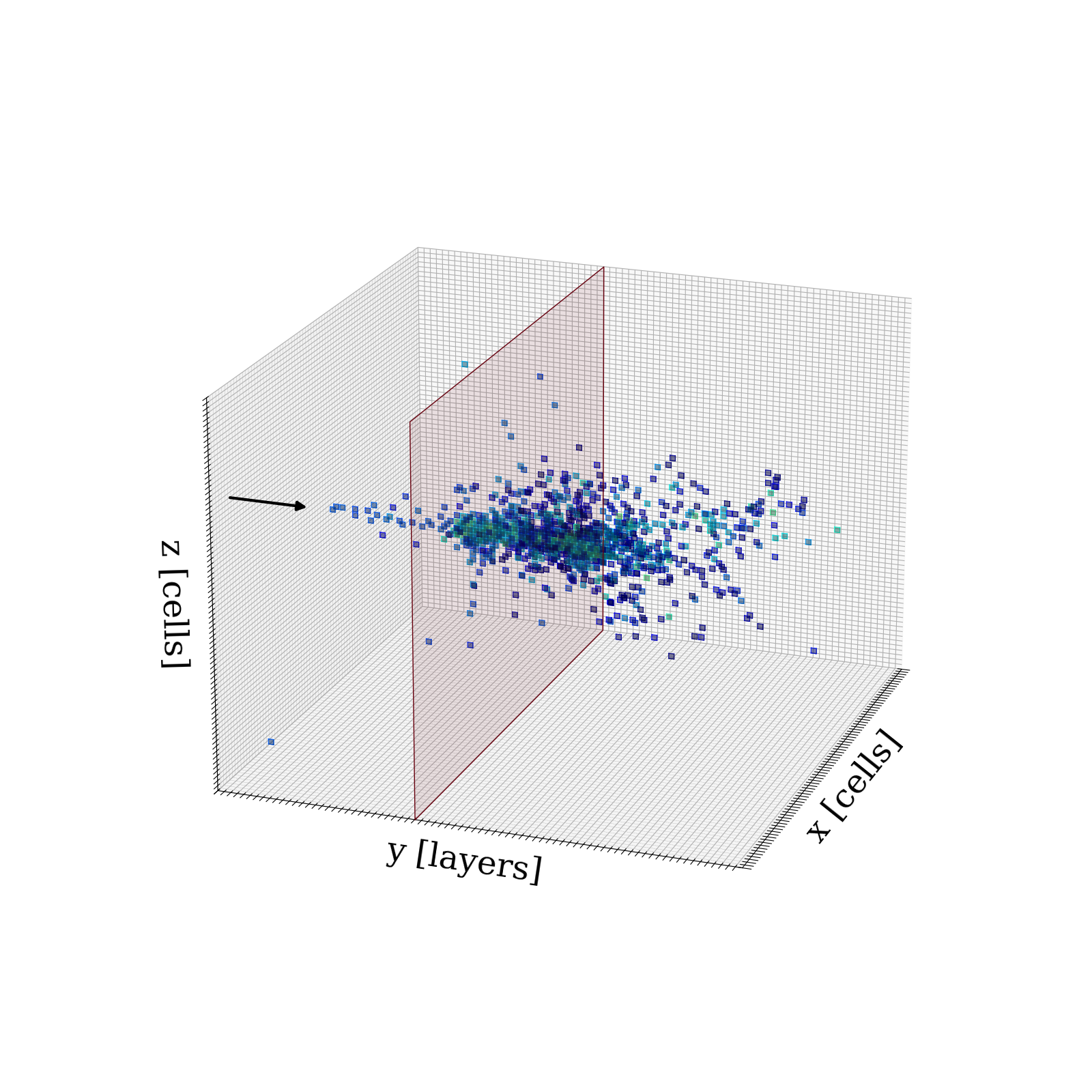}
    \hspace{-11mm}
    \includegraphics[width=55mm]{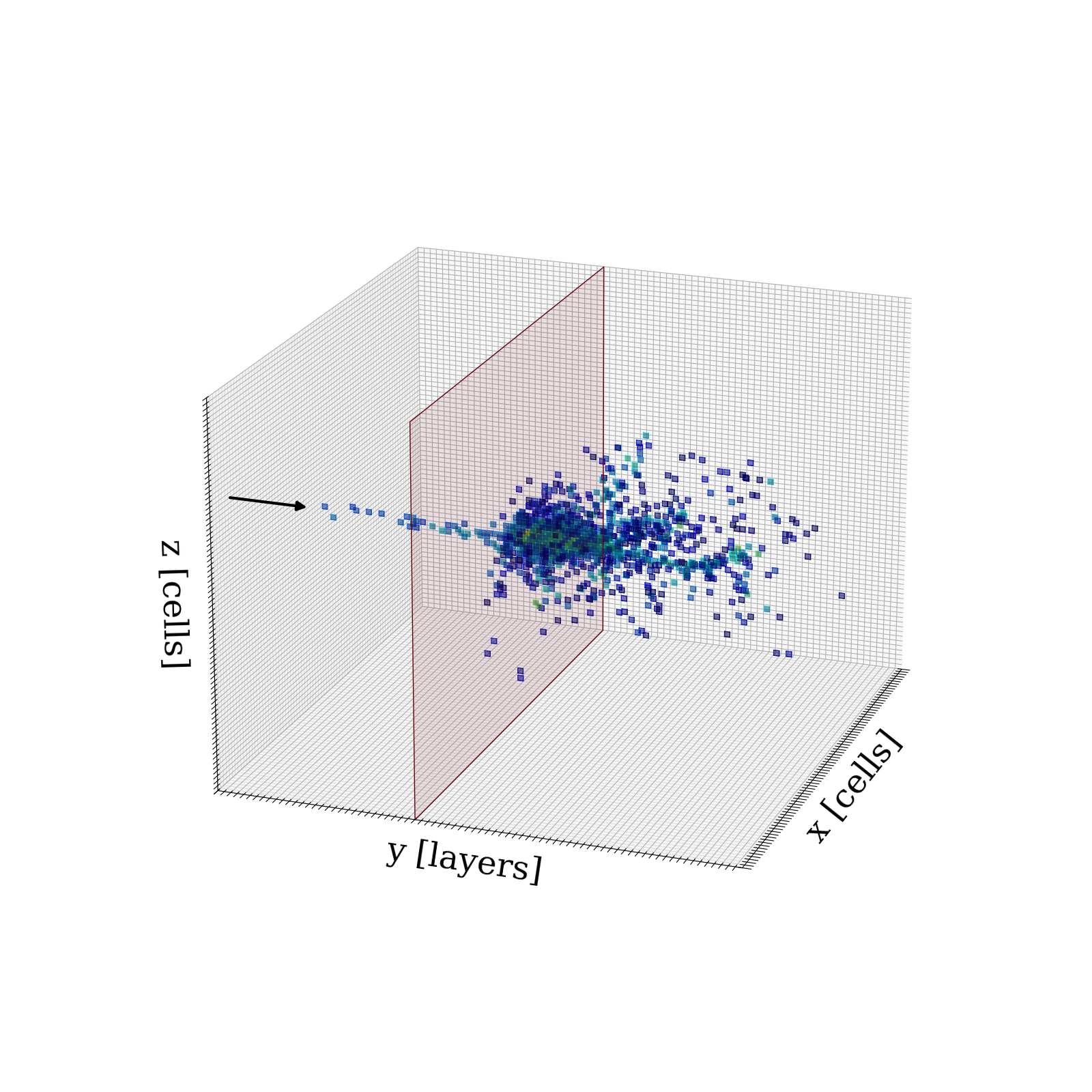}
    \caption{3D view of three 85~GeV \cc showers.
    The color represents the energy deposition in the cells. The red plane represents the division between ECal and HCal.}
    \label{fig:3d_85}
 \end{figure*}

Figures \ref{fig:3d_15}, \ref{fig:3d_50}, and \ref{fig:3d_85} show three examples of 15~GeV, 50~GeV and 85~GeV showers generated by \cc. 

\section{\texorpdfstring{\pointsFM}{PointCountFM} histograms}\label{sec:appC}

\begin{figure*}[htbp]
   \centering 
   \includegraphics[width=160mm]{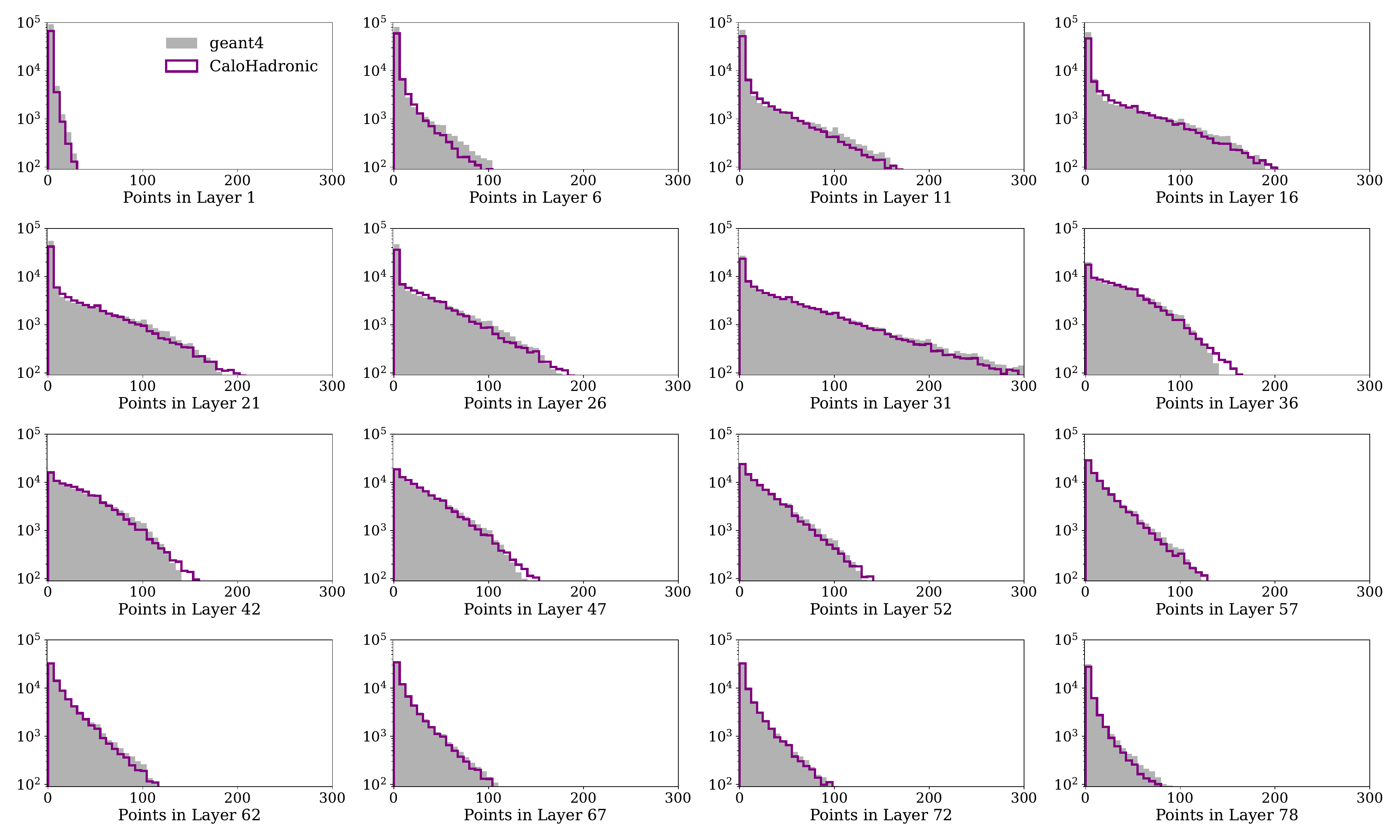}
   \caption{Histograms of the number of points per layer for Geant4 (gray) and \pointsFM (purple). All distributions are calculated for 100 000 showers with a uniform distribution of incident particle energies between 10 and 90 GeV.}
   \label{fig:PperL}
\end{figure*}

Figure \ref{fig:PperL} shows the generation of the points per layer in both ECal and HCal performed by the \pointsFM. One can see that the majority of points is usually around layer 30 (beginning of the HCal), which is where a shower most likely starts, while in the first and last layers the number of points decreases. 

A high quality modeling of these distributions is required as the points per layer output is used as conditioning to generate the hadronic showers and to further calibrate the number of points per layer in the shower. \pointsFM effectively ensures the required precision and accuracy for this process.

\section{\cc further histograms}\label{sec:appD}

\begin{figure*}[htbp]
   \centering 
   \hspace*{-2cm}
   \resizebox{\textwidth}{!}{
   \includegraphics[width=150mm]{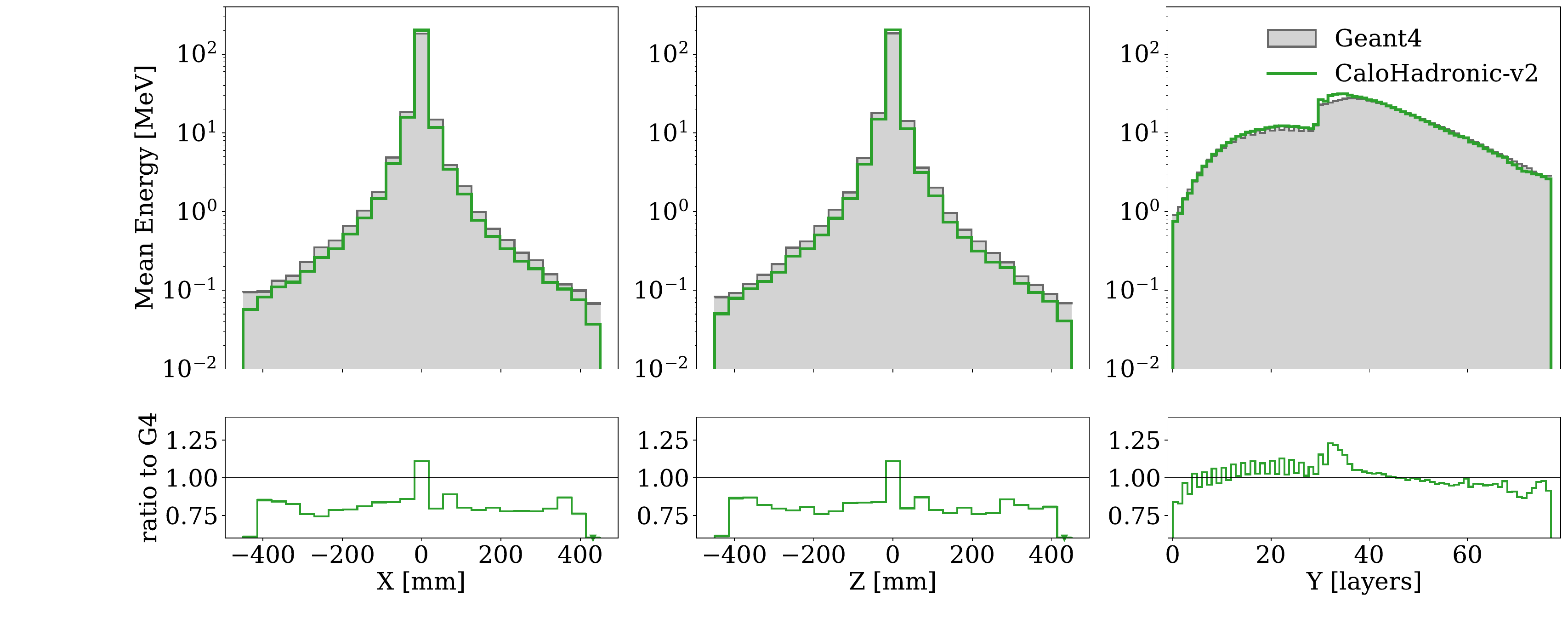}
   }
   \caption{Distribution of hit $x$ (left), $y$ (middle), and $z$ (right) positions calculated from $50 000$ showers. All distributions are calculated with a uniform distribution of incident particle energies 
   between 10 and 90 GeV. The error band corresponds to the statistical uncertainty 
   in each bin.}
   \label{fig:distxyz}
\end{figure*}

\begin{figure*}[htbp]
   \centering 
   \resizebox{\textwidth}{!}{
   \includegraphics[width=45mm]{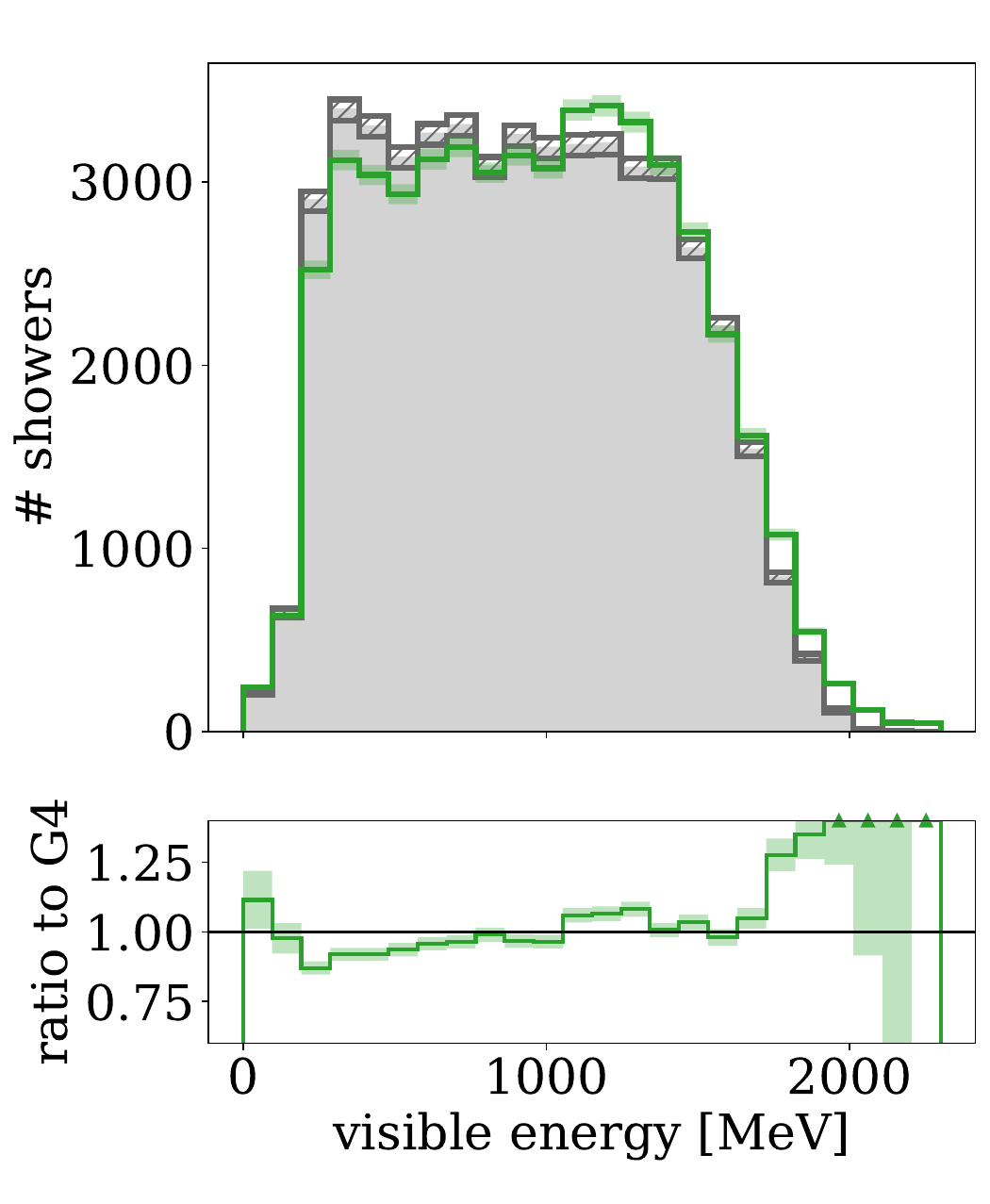}
   \includegraphics[width=45mm]{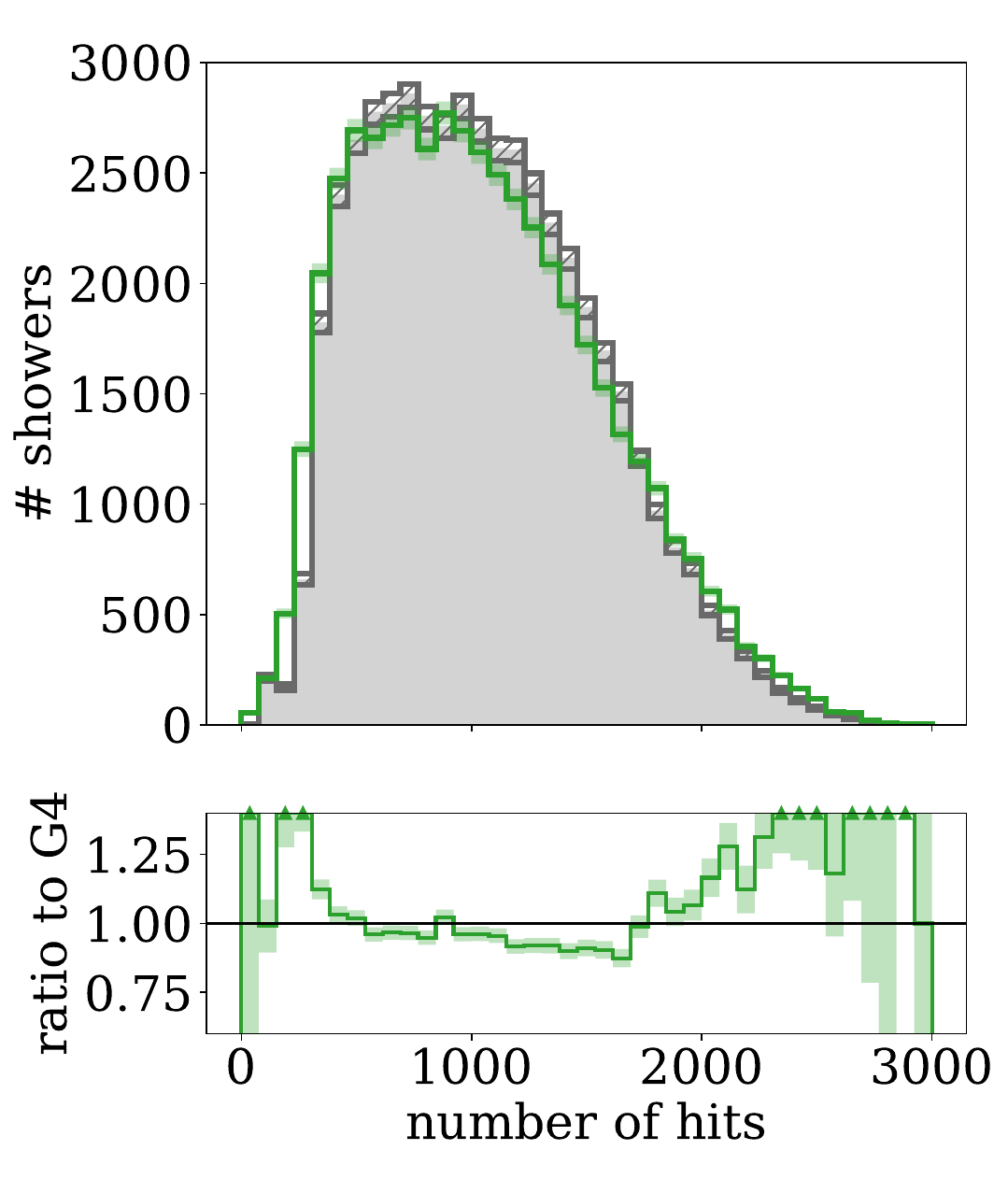} 
   \includegraphics[width=45mm]{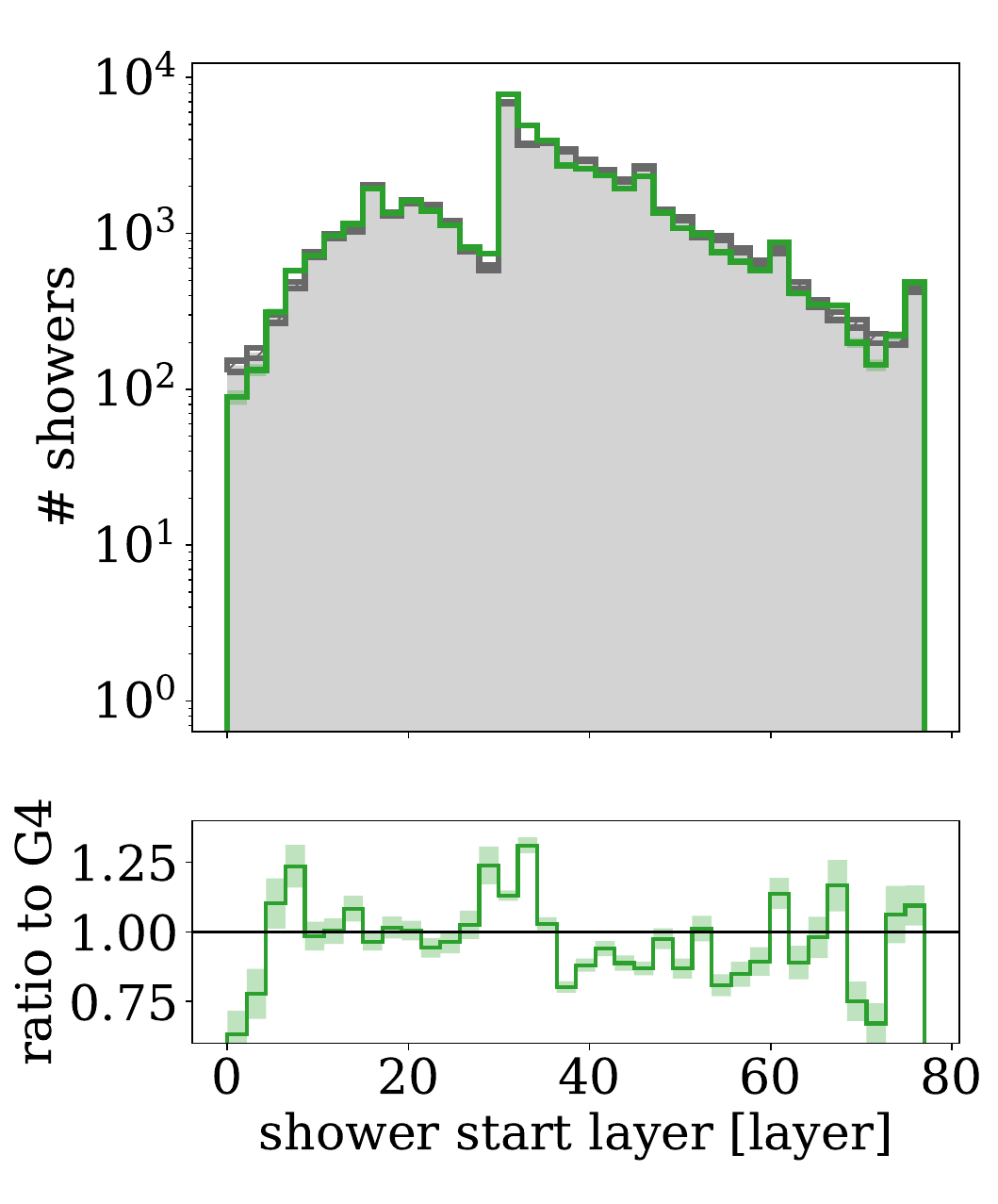}
   }
   \caption{Histograms of the total visible energy (left), 
   total number of hits (center), and shower start layer 
   (right) for Geant4 and \cc. All distributions are calculated with 
   $50 000$ events sampled from a uniform distribution of incident 
   particle energies between 10 and 90 GeV. The bottom panel provides 
   the ratio to Geant4. The error band corresponds to the 
   statistical uncertainty in each bin.}
   \label{fig:distapp}
\end{figure*}

Figures \ref{fig:distxyz} and \ref{fig:distapp} show additional \cc distributions for completeness. 

\section{Evaluation Scores}\label{app_eval}

We investigate the performance of \cc by calculating scores from 
high level calorimeter shower observables. The evaluation in this work relies on two metrics: the Wasserstein distance, which measures distributional similarity between features, and the quantile KL divergence, which compares the divergence between quantile distributions. \\ 
This provides a quantitative measure of the fidelity observed in the plots 
presented in Sec. \ref{subsec:pp}, allowing for an objective assessment 
rather than relying solely on a visual comparison of distributions. The evaluation scores are computed with 5 batches of $10,000$ showers.\\ 
The following observables are considered in order to calculate the one-dimensional 
scores: the number of hits (cells with energy depositions above 0.02~MeV threshold)
$N_{hits}$, the cell energy $E_{cell}$, the center of gravity in the 
$x$-, $y$-, and $z$-directions $cog_{x}$, $cog_{y}$ and $cog_{z}$, energy sum
$E_{sum}$, the longitudinal energy $E_{long}$, and the radial energy $E_{radial}$.

\begin{table}[htbp]
   \small 
   \centering
   \resizebox{\textwidth}{!}{ 
   \begin{tabular}{llllllllll}
   \toprule
   (x$ 10^{-2}$) & $cog_{x}$ & $cog_{y}$ & $cog_{z}$ & $Y_{start}$ & $E_{cell}$ & $E_{sum}$ & $E_{radial}$ & $E_{long}$ & $N_{hits}$ \\
   \midrule
   normalized WD  &  8.4 $\pm$ 0.8 & 3.8 $\pm$ 0.4 & 8.0 $\pm$ 0.7 & 5.2 $\pm$ 0.2 & 2.8 $\pm$ 0.4 & 9 $\pm$ 1 & 0.05 $\pm$ 0.02 & 76 $\pm$ 37 & 2.8 $\pm$ 0.4 \\
   \midrule
   quantile KL &  1.1 $\pm$ 0.3 & 0.6 $\pm$ 0.1& 0.9 $\pm$ 0.2 & 2.1 $\pm$ 0.2 & 0.3 $\pm$ 0.1 & 0.68 $\pm$ 0.07 & 0.2 $\pm$ 0.2 & 0.6 $\pm$ 0.7 & 0.3 $\pm$ 0.1 \\
   \bottomrule
   \end{tabular}
   }
   \caption{Wasserstein distances for several physics observables between generated and test data. The Wasserstein distances and the quantile KL are numerically evaluated using five batches of $10,000$ showers. Shown are the mean and the standard deviation over the five resulting values. }
   \label{table:eval} 
\end{table}
 
% \begin{figure*}
%    \centering 
%    \includegraphics[width=160mm]{figures/results/spider2.pdf}
%    \caption{Spider plot.}
%    \label{fig:spider}
% \end{figure*}
\subsection{Classifier Scores}\label{app:classifier}

We further compare the showers generated by the model to the Geant4 simulation by training a fully
connected high-level classifier to distinguish between model generated and Geant4 simulated showers. 
The 5 input shower observables are the three center of
gravity variables, the number of hits and energy ratio (total visible energy divided by the incident energy). 
150k Geant4 showers and 500k showers generated by \cc were used. A standard scaling of the inputs is applied. A $60\%$ , $20\%$, $20\%$ data split is applied for training set, validation and test set, respectively. The classifier is implemented as a fully connected neural network with three layers (containing
8, 8, 1 nodes respectively) with LeakyReLU \cite{Maas2013RectifierNI} activation functions. It is trained with the Adam optimizer \cite{kingma2017adammethodstochasticoptimization} for 100 epochs using a binary cross-entropy loss. The final model epoch is chosen based on the lowest validation loss.

To evaluate the performance of the classifier, we use two metrics: the area under the receiver operating characteristic curve (AUC) and the Jensen–Shannon divergence (JSD), both computed on the test set. These metrics are commonly used in the evaluation of generative models in high-energy physics, as demonstrated in previous works such as Ref. \cite{Buhmann:2023bwk, Buss:2024orz, PhysRevD.107.113003, Buckley:2023daw, Krause:2021wez, lopez-paz2017revisiting}.
The AUC reflects the classifier’s ability to distinguish between Geant4 and model-generated data. A perfect separation yields an AUC of 1.0, while an AUC of 0.5 indicates total confusion—implying that the generated data is indistinguishable from the simulated ones.

The JSD, on the other hand, is a symmetric and bounded measure of the difference between two probability distributions, often used to quantify how similar two datasets are. In this context, it measures the divergence between the classifier’s predicted probability distributions for simulated and generated samples. A JSD of 0 indicates perfect overlap (i.e., the distributions are identical), while higher values suggest greater divergence between the real and generated data.

To ensure robustness, the classifier was trained five times using different splits of the data into training, validation, and test sets.
In Tab. \ref{tab:classifier} we present the mean AUC and standard deviation of these five classifier trainings for \cc.
\begin{table}[ht]
   \small
   \centering
   \begin{tabular}{lcc}
      \toprule
      \textbf{Simulator} &\textbf{AUC} & \textbf{JSD}  \\
      \midrule 
      \cc  & 0.79 $\pm$ 0.03 & 0.2354 $\pm$ 0.0005 \\
       \bottomrule
   \end{tabular}
   \caption{Results of the high level classifier test. Model performance comparison with area under the receiver operating characteristic curve (AUC) score and Jensen–Shannon divergence (JSD). Shown are the mean and standard deviation over five random network initializations.}
   \label{tab:classifier}
\end{table}

\section{Resolution and Linearity}\label{app_res}

\begin{figure*}[htbp]
   \centering 
   \includegraphics[width=120mm]{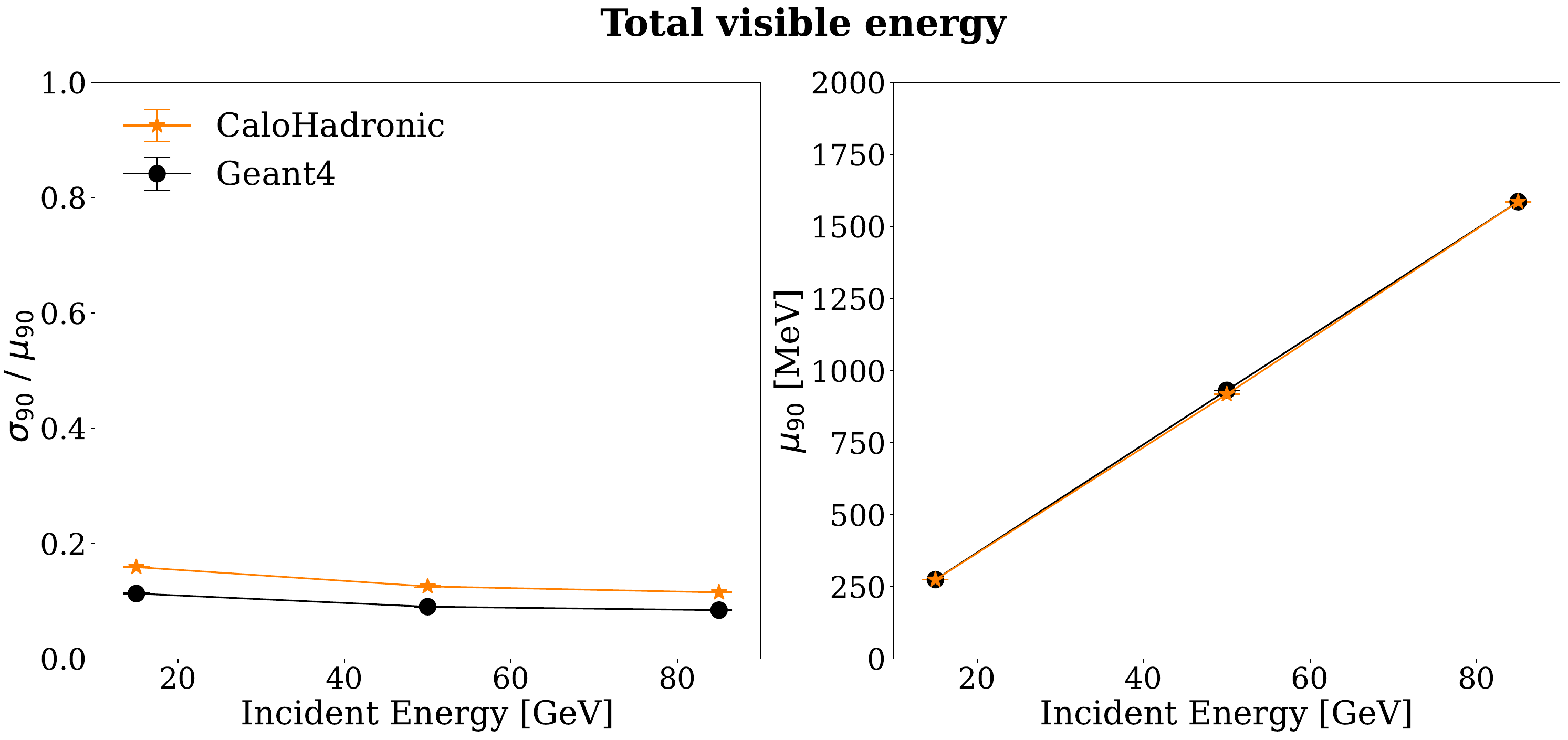}
   \caption{Relative resolution $\sigma_{90}$ / $\mu_{90}$ (on the left) and linearity $\mu_{90}$ (on the right) of the 90$\%$ core of the total visible energy per shower for Geant4 and \cc. The points are taken at incident energies of 15 GeV, 50 GeV, and 85 GeV. Geant4 showers are sampled within $\pm$ 1 MeV of the target energy; \cc showers are newly generated at the same incident energies.}
   \label{fig:resolution}
\end{figure*}

Figure \ref{fig:resolution} shows the relative resolution, computed by dividing the standard deviation by the mean ($\sigma$ / $\mu$) versus the incident energy, and the linearity, hence the mean ($\mu$) versus the incident energy, of the total visible energy per shower. Incident energies of 15 GeV, 50 GeV, and 85 GeV are shown. Geant4 showers are sampled within $\pm$ 1 MeV of the target energy; \cc showers are newly generated at the same incident energies. The mean and root-mean-square of the 90$\%$ core of these distributions, labeled $\mu_{90}$ and $\sigma_{90}$, is calculated for all energies. It should be noted that the resolution has not been corrected for the absorber thickness.

In figure \ref{fig:resolution}, in the linearity plots, the error on the mean is simply $\sigma / \sqrt{N}$. The relative resolution uncertainty was computed following \cite{rao1973linear}. Using the delta method approximation, the resulting formula is: 
\begin{equation}
    \mathrm{se}(s) \approx \frac{1}{2s} \sqrt{ \frac{1}{n} \left( \mu_4 - \frac{n - 3}{n - 1} \sigma^4 \right) }
\end{equation}
where $n$ is the sample size, $\mu_{4} \; = \; E[(X-\mu)^{4}]$, $\sigma^{4}$ is the square of the population variance and $s$ is the sample standard deviation. \\ 

Figure \ref{fig:resolution} shows that the resolution of \cc is slightly worse than Geant4, while the linearity plot shows good agreement between the two. Fitting a function of the form
$a + b / \sqrt{E_{0}} + c / E_{0}$ 
 to the incident energy ($E_{0}$) versus relative resolution ($\sigma_{90} / \mu_{90}$) distribution yields the following parameters for Geant4 (CaloHadronic): a: $0.063 \pm 0.003$ ($0.069 \pm 0.005$), b: $0.16 \pm 0.04$ ($0.40 \pm 0.06$), c: $0.16 \pm 0.04$ ($-0.2 \pm 0.1$). 
However, it should be noted that these values cannot be directly interpreted as resolution components, as they average over different materials and thicknesses.

\end{document}